\documentclass[onecolumn,10pt,conference]{IEEEtran}

\usepackage{graphicx}
\usepackage{amssymb}
\usepackage{amsmath}
\usepackage{psfrag}

\newtheorem{theorem}{Theorem}
\newtheorem{cor}{Corollary}
\newtheorem{lem}{Lemma}

\newtheorem{defn}{Definition}

\newcommand{\brkt}[1]{\langle#1\rangle}

\newcommand{\bx}{\mathbf{x}}
\newcommand{\by}{\mathbf{y}}
\newcommand{\bz}{\mathbf{z}}
\newcommand{\bs}{\mathbf{s}}
\newcommand{\bb}{\mathbf{b}}
\newcommand{\bd}{\mathbf{d}}
\newcommand{\bB}{\mathbf{\mathbf{B}}}

\newcommand{\bZ}{\mathbf{Z}}

\newcommand{\bC}{\mathbf{C}}
\newcommand{\bV}{\mathbf{V}}

\newcommand{\Wbicm}{\widetilde{W}}
\newcommand{\WbicmSym}{\overline{W}}
\newcommand{\PBICM}{\textrm{PBICM}}
\newcommand{\BICM}{\textrm{BICM}}

\newcommand{\nhs}{\hspace{-.04in}}

\newcommand{\eps}{\varepsilon}

\newcommand{\Y}{\mathcal{Y}}
\newcommand{\X}{\mathcal{X}}
\newcommand{\C}{\mathcal{C}}
\newcommand{\bc}{\mathbf{c}}

\newcommand{\E}{\mathbf{E}}

\newcommand{\Err}{\mathcal{E}}

\newcommand{\EE}{\mathbb{E}}
\newcommand{\VAR}{V\!AR}
\renewcommand{\S}{\mathcal{S}}
\newcommand{\ra}{\rightarrow}

\long\def\symbolfootnote[#1]#2{\begingroup%
\def\thefootnote{\fnsymbol{footnote}}\footnote[#1]{#2}\endgroup}
\begin{document}

\title{ Parallel Bit Interleaved Coded Modulation}

\author{Amir Ingber$^\dagger$  and Meir Feder\\
Department of EE-Systems, Tel Aviv University\\
Tel Aviv 69978, ISRAEL\\
\{ingber, meir\}@eng.tau.ac.il}

\maketitle

\symbolfootnote[0]{ $^\dagger$
A. Ingber is supported by the Adams Fellowship Program of the Israel Academy of Sciences and Humanities. }

\begin{abstract}

A new variant of bit interleaved coded modulation (BICM) is proposed. In the new scheme, called \emph{parallel BICM}, $L$ identical binary codes are used in parallel using a mapper, a newly proposed finite-length interleaver and a binary dither signal. As opposed to previous approaches, the scheme does not rely on any assumptions of an ideal, infinite-length interleaver.
Over a memoryless channel, the new scheme is proven to be equivalent to a \emph{binary} memoryless channel. Therefore the scheme enables one to easily design coded modulation schemes using a simple binary code that was designed for that binary channel. The overall performance of the coded modulation scheme is analytically evaluated based on the performance of the binary code over the binary channel.
The new scheme is analyzed from an information theoretic viewpoint, where the capacity, error exponent and channel dispersion are considered. The capacity of the scheme is identical to the BICM capacity. The error exponent of the scheme is numerically compared to a recently proposed mismatched-decoding exponent analysis of BICM.
\end{abstract}

\section{Introduction}

Bit interleaved coded modulation (BICM) is a pragmatic approach for coded modulation \cite{ZehaviBICM94}. It enables the construction of nonbinary communication schemes from binary codes by using a long bit interleaver that separates the coding and the modulation. BICM has drawn much attention in recent years, because of its efficiency for wireless and fading channels.

The information-theoretic properties of BICM were first studied by Caire et. al. in \cite{Caire_BICM_98}. BICM was modeled as a binary channel with a random state that is known at the receiver. The state determines how the input bit is mapped to the channel, along with the other bits that are assumed to be random. Under the assumption of an infinite-length, ideal interleaver, the BICM scheme is modeled by parallel uses of independent instances of this binary channel. This model is referred to as the \emph{independent parallel channel model}.

Using this model the capacity of the BICM scheme could be calculated. It was further shown that BICM suffers from a gap from the full channel capacity, and that when Gray mapping is used this gap is generally small. In \cite{Caire_BICM_98}, methods for evaluating the error probability of BICM were proposed, which rely on the properties of the specific binary codes that were used (e.g. Hamming weight of error events).

A basic information-theoretic quantity other than the channel capacity is the error exponent \cite{GallagerInfoTheoryBook}, which quantifies the speed at which the error probability decreases to zero with the block length $n$. Another tool for evaluating the performance at finite block length is the channel dispersion, which was presented in 1962 \cite{Strassen62_Asymptotische} and was given more attention only in recent years \cite{PolyanskiyPV09_GaussianDispersion}, \cite{PolyanskiyPVFiniteLength10}. It would therefore be interesting to analyze BICM at finite block length from the information-theoretic viewpoint.

Several attempts have been made to provide error exponent results for BICM.

In their work on multilevel codes, Wachsmann et. al. \cite{WachsmannFischerHuber_MLC_99} have considered the random coding error exponent of BICM, by relying on the independent parallel channels model. However, there were several flaws in the derivation:
\begin{itemize}
  \item The independent parallel channels model is justified by an infinite-length interleaver. Therefore it might be problematic to use its properties for evaluating the finite length performance of the BICM scheme. In the current paper we address this point and propose a scheme with a finite-length interleaver for that purpose.
  \item There was a technical flaw in the derivation, which resulted in an inaccurate expression for the random coding error exponent. We discuss this point in detail in Theorem \ref{thm:WbicmSymExp}.
  \item As noticed in \cite{MartinezFabregas09_BICM_mismatched}, the error exponent result obtained in \cite{WachsmannFischerHuber_MLC_99} sometimes may even exceed that of unconstrained coding over the channel (called in \cite{MartinezFabregas09_BICM_mismatched} the ``coded modulation exponent''). We therefore agree with \cite{MartinezFabregas09_BICM_mismatched} in the claim that ``the independent parallel channel model fails to capture the statistics of the channel''. However, by properly designing the communication scheme the model can become valid in a rigorous way, as we show in Theorem \ref{thm:equivalence}.
\end{itemize}

In \cite{MartinezFabregas09_BICM_mismatched} (see also \cite{FoundadationsAndTrendsBICM2008}), Martinez et al. have considered the BICM decoder as a mismatched decoder, which has access only to the log-likelihood values (LLR) of each bit, where the LLR calculation assumes that the other bits are random, independent and equiprobable (as in the classical BICM scheme \cite{Caire_BICM_98}). Using results from mismatched decoding, they presented the generalized error exponent and the generalized mutual information, and pinpointed the loss of BICM that incurs from using the mismatched LLRs. Note that when a binary code of length $n$ is used, the scheme requires only $n/L$ channel uses. While this result is valid for any block size and any interleaver length, achieving this error exponent in practice requires complex code design. For example, one cannot design a good binary code for a binary memoryless channel and have any guarantee that the BICM scheme will perform well with that code. In fact, the code design for this scheme requires taking into account the memory within the levels, or equivalently, nonbinary codes, which is what we wish to avoid when choosing BICM.

On the theoretical side, another drawback of existing approaches is the lack of converse results (for either capacity or error exponent). The initial discussion of BICM information theory in \cite{Caire_BICM_98} assumes the model of independent channels, and any converse result based on this model must assume that an infinite, ideal, interleaver. Therefore the converse results (such as upper bound on the achievable rate with BICM) do not hold for finite-length interleavers. The authors in \cite{MartinezFabregas09_BICM_mismatched} provide no converse results for their model.

In this paper we propose the \emph{parallel BICM} (PBICM) scheme, which has the following properties. First, the scheme includes an explicit, \emph{finite length} interleaver. Second, in order to attain good performance on any memoryless channel, PBICM allows one to design a binary code for a binary memoryless channel, and guarantees good performance on the nonbinary channel. Third, because the scheme does not rely on the use of an infinite-length interleaver, the error exponent and the dispersion of the scheme can be calculated (both achievability and converse results) as means to evaluate the PBICM performance at finite block length.

The comparison between PBICM and the mismatched decoding approach \cite{MartinezFabregas09_BICM_mismatched} should be done with care. With PBICM, when the binary codeword length is $n$ the scheme requires $n$ channel uses. Therefore when the latency kept equal for both schemes, PBICM uses a codeword length that is $L$ times shorter than the codeword used in the mismatched decoder. A fair comparison would be to fix the binary codeword length $n$ for both schemes, resulting in different latency, but equal decoder complexity.

The results presented in the paper are summarized as follows:

\begin{itemize}
  \item The PBICM communication framework is presented. Over a memoryless channel, it is shown to be equivalent to a \emph{binary} memoryless channel (Theorem \ref{thm:equivalence}).
  \item In Theorem \ref{thm:PBICMcapacity}, the capacity of PBICM is shown to be equal to the BICM capacity, as calculated in \cite{Caire_BICM_98}.
  \item PBICM is analyzed at finite block length. The error exponent of PBICM is defined and bounded by error exponent bounds of the underlying binary channel (Theorems \ref{thm:PBICMerrexp} and \ref{thm:WbicmSymExp}).
  \item The PBICM dispersion is defined as an alternative measure for finite-length performance. It is calculated by the dispersion of the underlying binary channel (Theorems \ref{thm:PBICMdispersion} and \ref{thm:WbicmDisp}).
  \item The error exponent of PBICM is numerically compared to the mismatched-decoding error exponent of BICM \cite{MartinezFabregas09_BICM_mismatched}. The additive white Gaussian noise (AWGN) channel and the Rayleigh fading channel are considered. When the latency of both schemes is equal, the mismatched-decoding is generally better. However, when the complexity is equal (or where the codeword length of the underlying binary code is equal), the PBICM exponent is better in many cases.
\end{itemize}

The paper is organized as follows.

In Section \ref{sec:BICMmodel} we review the classical BICM model and its properties, under the assumption of an infinite-length, ideal interleaver.
In Section \ref{sec:PBICM} the parallel BICM scheme is presented and the equivalence to a memoryless binary channel is established.
In Section \ref{sec:PBICM_IT} parallel BICM is studied from an information-theoretical viewpoint.
Numerical examples and summary follow in Sections \ref{sec:numerics} and \ref{sec:summary} respectively.

\section{The BICM Communication Model}\label{sec:BICMmodel}

Notation: letters in bold ($\bx, \by$...) denote row vectors, capital letters ($X, Y$...) denote random variables, and tilde denotes interleaved signals ($\tilde{\bb}, \tilde{\bz}$). $P_X(x)$ denotes the probability that the random variable (RV) $X$ will get the value $x$, and similarly $P_{Y|X}(y|x)$ denotes the probability $Y$ will get the value $y$ given that the RV $X$ is equal to $x$. $\EE[\cdot]$ denotes statistical expectation. $\log$ means $\log_2$.

\subsection{Channel model}
Let $W$ denote a memoryless channel with input and output alphabets $\X$ and $\Y$ respectively. The transition probabilities are defined by $W(y|x)$ for $y \in \Y$ and $x \in \X$. We assume that $\|\X\| = 2^L$. We consider equiprobable signaling only over the channel $W$.

An $(n,R)$ code $\C \subseteq \X^n$ is a set of $M=2^{nR}$ codewords $\bc \in \X^n$. The encoder wishes to convey one of $M$ equiprobable
messages. The error probability of interest shall be the codeword error probability. An $(n,R)$ code with codeword error probability $p_e$ will sometimes be called an $(n,R,p_e)$ code.

\subsection{Classical BICM encoding and decoding}
\begin{figure}\begin{center}
    \psfrag{&m}{$\!\!\!\!\!\!\!\!\!\!\!  m_1, \! m_2..$}
    \psfrag{&binary}{$\! \! \! \! \! \! \! \begin{array}{c}
                        \textrm{\small Binary}\\
                        \textrm{\small encoder}
                  \end{array}$ }
    \psfrag{&b}{$\!\!\!\!\bb_1\!,\bb_2..$}
    \psfrag{&pi}{$\pi$}
    \psfrag{&u}{$\mu$}
    \psfrag{&xt}{$\!\!\bx$}
    \psfrag{&t}{$\tilde{\bb}$}

  \includegraphics[width=3.5in]{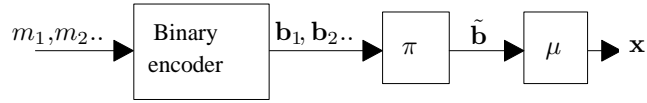}\\
  \caption{BICM encoding process}\label{fig:BICMenc}
\end{center}
\end{figure}

In BICM, a binary code is used to encode information messages $[m_1,m_2,...]$ into binary codewords $[\bb_1,\bb_2,...]$. The binary codewords are then interleaved using a long interleaver $\pi(\cdot)$, which applies a permutation on the coded bits. The interleaved bit stream $\tilde{\bb}$ is partitioned into groups of $L$ consecutive bits and inserted into a mapper $\mu:\{0,1\}^L \to \X$. The mapper output, denoted $\bx$, is fed into the channel. The encoding process is described in Figure \ref{fig:BICMenc}.

\begin{figure}\begin{center}
    \psfrag{&mh}{$\!\!\!\!\!\!\!  \hat{m}_1, \! \hat{m}_2..$}
    \psfrag{&binary}{$\! \! \! \! \! \! \! \begin{array}{c}
                        \textrm{\small Binary}\\
                        \textrm{\small dec.}
                  \end{array}$ }
    \psfrag{&llr}{$\! \! \! \! \! \! \! \begin{array}{c}
                        \textrm{\small LLR}\\
                        \textrm{\small calc.}
                  \end{array}$ }
    \psfrag{&z}{$\!\!\!\!\!\!\!\!\!\bz_1\!,\bz_2..$}
    \psfrag{&pi}{$\!\!\pi^{\!-\!1}$}
    \psfrag{&u}{$\!\!W$}
    \psfrag{&xt}{$\!\!\bx$}
    \psfrag{&y}{$\by$}
    \psfrag{&zt}{$\tilde{\bz}$}

  \includegraphics[width=3.5in]{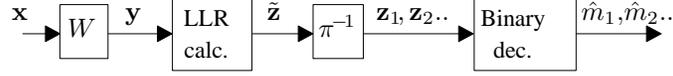}\\
  \caption{BICM decoding process}\label{fig:BICMdec}
\end{center}\end{figure}

The decoding process of BICM proceeds as follows. The channel output $\by$ is fed into a bit metric calculator, which calculates the log-likelihood ratio (LLR) of each input bit $b$ given the corresponding output sample $y$ ($L$ LLR values for each output sample). These LLR values  (or bit metrics) denoted  $\tilde{\bz}$ are de-interleaved and partitioned into bit metrics $[\bz_1,\bz_2,...]$ that correspond to the binary input codewords. Finally, the binary decoder decodes the messages $[\hat{m}_1,\hat{m}_2,...]$ from $[\bz_1,\bz_2,...]$. The decoding process is described in Figure \ref{fig:BICMdec}.

The LLR of the $j^{th}$ bit in a symbol given the output value $y$ is calculated as follows:
\begin{equation}\label{eqn:LLRbicm}
    LLR_j(y) \triangleq \log \frac{P_{Y|B_j}(y|0)}{P_{Y|B_j}(y|1)},
\end{equation}
where $P_{Y|B_j}(y|b)$ is the conditional probability of the channel output getting the value $y$ given that the $j^{th}$ bit at the mapper input was $b$, and the other $(L-1)$ bits are equiprobable independent binary random variables (RVs).

\subsection{Classical BICM analysis: ideal interleaving}

\begin{figure}\begin{center}
    \psfrag{&m}{$\ \mu$}
    \psfrag{&b1}{$B_1$}
    \psfrag{&b2}{$B_2$}
    \psfrag{&bj}{$\quad B_i$}
    \psfrag{&bL}{$B_L$}
    \psfrag{&x}{$\ X$}
    \psfrag{&y}{$\!Y$}
    \psfrag{&w}{$W$}
    \psfrag{&wj}{$\ \ W_i$}
    \psfrag{&vd}{\ :}

  \includegraphics[width=3.5in]{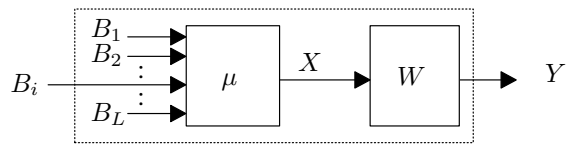}\\
  \caption{The binary channel $W_i$. The bits $\{B_j\}_{j \neq i}$ are equiprobable independent RVs.}\label{fig:Wi}
\end{center}
\end{figure}

In classical BICM (e.g. \cite{Caire_BICM_98}) the LLR calculation is motivated by the assumption of a very long (\emph{ideal}) interleaver $\pi$, so the coded bits go through essentially \emph{independent} channels. These binary channels are defined as follows:

\begin{defn}
Let $W_i$ be a binary channel with transition probability
\begin{eqnarray}
  W_i(y|b)\hspace{-.05in} &\triangleq&\hspace{-.05in} \EE \left[W(y|X=\mu(B_1,...,B_L)) | B_i = b\right] \\
   &=& \hspace{-.05in}\frac{1}{2^{L-1}}\sum_{\underset{b_i = b}{b_j;\ i \neq j}} W(y|\mu(b_1,...,b_L)).
\end{eqnarray}

The channel $W_i(y|b_i)$ can be thought of as the original channel $W$ where the input is $x = \mu(b_1...b_L)$, where the bits $\left\{b_j \right\}_{j\neq i}$ are equiprobable independent RVs (see Fig. \ref{fig:Wi}).
\end{defn}

\begin{figure}\begin{center}
    \psfrag{&b}{$\quad \ \ B$}
    \psfrag{&S}{\!\!\!$S$: random state}
    \psfrag{&s}{$S$}
    \psfrag{&x}{$\ X$}
    \psfrag{&y}{$\!Y$}
    \psfrag{&ws}{$W_s$}

  \includegraphics[width=2.6in]{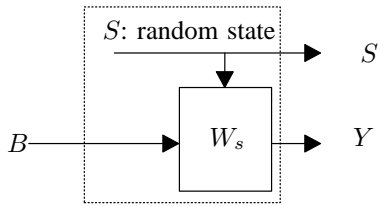}\\
  \caption{The binary channel $\Wbicm$. The random state $S$ is known at the receiver.}\label{fig:Wbicm}
\end{center}
\end{figure}

In \cite{Caire_BICM_98}, Caire et al. have proposed the following channel model for BICM called the \emph{independent parallel channel model}. In this model the channel has a binary input $b$. A channel state $s$ is selected at random from $\S \triangleq \{1,...,L\}$ with equal probability (and independently of $b$). Given a state $s$, the input bit $b$ is fed into the channel $W_s$. The channel outputs are the state $s$ and the output $y$ of the channel $W_s$. The channel, denoted by $\Wbicm$, is depicted in Figure \ref{fig:Wbicm}.

The transition probability function of $\Wbicm$ is given by
\begin{eqnarray}
  \Wbicm(y,s|b) &=& P_{Y,S|B}(y,s|b) \nonumber\\
  &=& P_{Y|S,B}(y|s,b) P_S(s)  \nonumber\\
  &=& \frac{1}{L}W_s(y,b).\label{eqn:WbicmProbability}
\end{eqnarray}

Note that both outputs can be combined into a single output, the LLR, which is a sufficient statistic for optimal decoding over any binary-input channel. The LLR calculation for the channel $\Wbicm$ is given by
\begin{equation}\label{eqn:LLRWbicm}
    LLR_{\Wbicm}(y,s) = LLR_s(y),
\end{equation}
where $LLR_s$ is given in (\ref{eqn:LLRbicm}).

Therefore the independent parallel channel model transforms the original nonbinary channel $W$ to a simple, memoryless channel. Using an infinite-length interleaver and a binary code that was designed for the simple binary channel $\Wbicm$, reliable communication for the original channel $W$ can be attained.

Let $\bC(\cdot )$ denote the Shannon capacity of a channel (with equiprobable input).

\begin{lem}[following \cite{Caire_BICM_98}]\label{lem:CBICM}
Let $\bC^{\BICM}(W)$ denote the capacity of the channel $W$ with BICM, a given mapping $\mu(\cdot)$ and an infinite-length interleaver (according to the independent parallel channel model). Denote by $\bC(W_s)$ the capacity of the channel $W_s$. Then
\begin{equation}\label{eqn:CBICM}
    \bC^{\BICM}(W) = \sum_{s=1}^L \bC(W_s).
\end{equation}
\begin{proof}
Since the independent parallel channel model assumes $L$ independent uses of the channel $\Wbicm$, we get that $\bC^{\BICM}(W) = L\cdot \bC(\Wbicm)$. The capacity of $\Wbicm$ is given by
\begin{eqnarray}
  \bC\left(\Wbicm\right) &=& I(B;Y,S) = I(B;Y|S)\nonumber\\
   &=& \EE_S I(B;Y|S=s) = \EE_S  \bC(W_s) \nonumber\\
   &=& \frac{1}{L} \sum_{s=1}^L \bC(W_s).\label{eqn:BICMcapacity}
\end{eqnarray}
\end{proof}
\end{lem}

It is known that $\bC^{\BICM}(W)$  is generally smaller than the full channel capacity $\bC(W)$, as opposed to other schemes, most notably multilevel coding and multistage decoding (MLC-MSD) \cite{WachsmannFischerHuber_MLC_99}, in which $\bC(W)$ can be achieved. However, for Gray mapping the gap is small and can sometimes be tolerated. For example, for 8-PSK signaling over the AWGN channel with SNR = 5dB, $\bC(W)=1.86bit$ where $\bC^{\BICM}(W)=1.84bit$.

\section{The Parallel BICM Scheme}\label{sec:PBICM}

In this section we propose an explicit BICM-type communication scheme which we call \emph{parallel BICM} (PBICM), which allows the usage of binary codes on nonbinary channels at finite blocklength. The main features of the scheme include the following:
\begin{itemize}
  \item Binary codewords are used \emph{in parallel} to construct a codeword that enters the channel.
  \item A new finite-length interleaver.
  \item A random binary signal (binary dither) that is added to the binary codewords.
\end{itemize}
With the proposed scheme, we rigorously show how the original channel $W$ relates to the channel $\Wbicm$, thus allowing exact analysis and design of codes at finite block lengths.

\subsection{Interleaver Design}

We wish to design a finite length interleaver, where:

\begin{itemize}
  \item The length of the interleaver is minimal,
  \item The interleaver should be as simple as possible,
  \item The binary codewords will go through a binary memoryless channel.
\end{itemize}

In order for the binary codewords to experience a memoryless channel, each binary codeword must be spread over $n$ channel uses of $W$, so the interleaver output length cannot be less than $n$ channel uses. The newly proposed interleaver has of output length of exactly $n$, which satisfies the above requirements.

Let $ENC$ and $DEC$ be an encoder-decoder pair for a binary code. Let $\bb_1,...,\bb_L$ be $L$ consecutive codewords from the output of $ENC$, bunched together to a matrix $\bB$:
\begin{equation}\label{eqn:interleaver}
\bB =
\left(
  \begin{array}{c}
  \bb_1 \\
  \vdots \\
  \bb_L \\
  \end{array}
\right)
=
\left(
  \begin{array}{cccc}
  b_{11} & \hdots & b_{1n} \\
  \vdots & b_{lk} & \vdots \\
  b_{L1} & \hdots & b_{Ln} \\
  \end{array}
\right).
\end{equation}

Let $\bs$ be a vector of i.i.d. random states drawn from $\S^n=\{1..L\}^n$. $\bs$ shall be the interleaving signal. Each column in $\mathbf{B}$ shall be shifted cyclically by the corresponding element $s_k$, so the interleaved signal $\tilde{\bB}$ is defined as
\begin{equation*}
\tilde{\bB} =
\left(
  \begin{array}{cccc}
  b_{(1+s_1)_L1} & \hdots & b_{(1+s_n)_Ln} \\
  \vdots & b_{(l+s_k)_Lk} & \vdots \\
  b_{(L+s_1)_L1} & \hdots & b_{(1+s_n)_Ln} \\
  \end{array}
\right),
\end{equation*}
where $(\xi)_L \triangleq (\xi \textrm{ modulo } L)+1$.

Each column vector of interleaved signal $\tilde{\bB}$ is mapped to a single channel symbol:
\begin{equation}
    x_k = \mu(b_{(1+s_k)_Lk},\hdots,b_{(L+s_k)_Lk}),
\end{equation}
and we call $\bx=[x_1,...,x_n]$ the \emph{channel codeword}.

At the decoder an LLR value is calculated for every bit $b$ in $\tilde \bB$ from $\by$. The LLR values are denoted by $\tilde{\bZ}$. We assume that $\bs$ is known at the decoder (utilizing common randomness), therefore  the de-interleaving operation is simply sorting back the columns of $\tilde\bZ$ according to $\bs$ by reversing the modulo operation. The de-interleaver output is a vector of LLR values $\bz$ for each transmitted codeword $\bb$, according to (\ref{eqn:LLRbicm}). Each codeword is decoded independently by $DEC$.

\subsection{Binary dither}

\begin{figure}\begin{center}
    \psfrag{&b1}{$\bb_1$}
    \psfrag{&bL}{$\bb_L$}
    \psfrag{&m1}{$m_1$}
    \psfrag{&mL}{$m_L$}
    \psfrag{&E1}{$\!ENC_1$}
    \psfrag{&EL}{$\!ENC_L$}
    \psfrag{&vd}{$\vdots$}
    \psfrag{&s}{$\bs$}
    \psfrag{&pi}{$\ \pi$}
    \psfrag{&u}{$\mu$}
    \psfrag{&xt}{$\!\!\bx$}
    \psfrag{&Bt}{$\tilde \bB$}
  \includegraphics[width=3.5in]{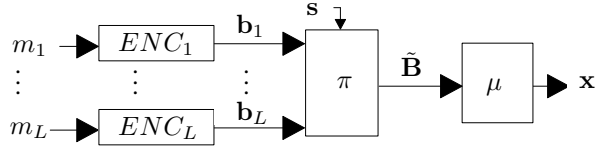}\\
  \caption{Interleaving scheme viewed as parallel encoders}\label{fig:BICMencoderNew}
\end{center}\end{figure}

\begin{figure}\begin{center}
    \psfrag{&y}{$\by$}
    \psfrag{&llr}{$\! \! \! \! \! \! \! \begin{array}{c}
                        \textrm{\small LLR}\\
                        \textrm{\small calc.}
                  \end{array}$ }
    \psfrag{&Zt}{$\tilde{\bZ}$}
    \psfrag{&s}{$\bs$}
    \psfrag{&pi}{$\pi^{-1}$}

    \psfrag{&z1}{$\bz_1$}
    \psfrag{&zL}{$\bz_L$}
    \psfrag{&D1}{$\! DEC_1$}
    \psfrag{&DL}{$\! DEC_L$}
    \psfrag{&vd}{$\vdots$}
    \psfrag{&m1}{$\hat m_1$}
    \psfrag{&mL}{$\hat m_L$}
    \psfrag{&x}{$\!\!\bx$}

  \includegraphics[width=3.5in]{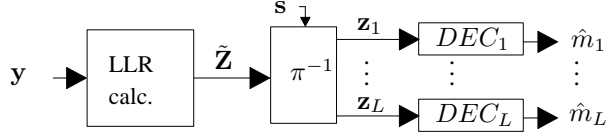}\\
  \caption{De-interleaving scheme viewed as parallel decoders}\label{fig:BICMdecoderNew}
\end{center}\end{figure}

Since the decoder decodes each binary codeword independently, the communication scheme employing the above interleaver
can be viewed as as set of parallel encoder-decoder pairs, which we denote by $ENC_1,...,ENC_L$ and $DEC_1,...,DEC_L$ (see Figures \ref{fig:BICMencoderNew} and \ref{fig:BICMdecoderNew}). We do not assume any independence between the effective channels between each encoder-decoder pair.

Consider the first encoder-decoder pair, $ENC_1$ and $DEC_1$. Since the input of $DEC_1$ depends on the codewords
transmitted by $ENC_2$,...,$ENC_L$, the channel between $ENC_1$ and $DEC_1$ is not strictly memoryless. If, somehow, the decoders
$DEC_2$,...,$DEC_L$ were forced to send i.i.d. equiprobable binary codewords, then the channel between $ENC_1$ and $DEC_1$ would be exactly
the channel $\Wbicm$ (which is a binary memoryless channel) with the accurate LLR calculation (\ref{eqn:LLRbicm}).

In order to achieve the goal of $L$ binary memoryless channels between each encoder-decoder pair simultaneously,
we add a binary dither - an i.i.d. equiprobable binary signal - to each encoder-decoder pair as follows.

\begin{figure}\begin{center}
    \psfrag{&b1}{$\bb_1$}
    \psfrag{&bL}{$\bb_L$}
    \psfrag{&b1p}{$\bb_1'$}
    \psfrag{&bLp}{$\bb_L'$}
    \psfrag{&m1}{$m_1$}
    \psfrag{&mL}{$m_L$}
    \psfrag{&d1}{$\bd_1$}
    \psfrag{&dL}{$\bd_L$}
    \psfrag{&p}{$\!_{^+}$}
    \psfrag{&E1}{$\!ENC_1$}
    \psfrag{&EL}{$\!ENC_L$}
    \psfrag{&vd}{$\vdots$}
    \psfrag{&s}{$\bs$}
    \psfrag{&pi}{$\ \pi$}
    \psfrag{&u}{$\mu$}
    \psfrag{&xt}{$\!\!\bx$}
    \psfrag{&Bt}{$\tilde \bB$}
  \includegraphics[width=4in]{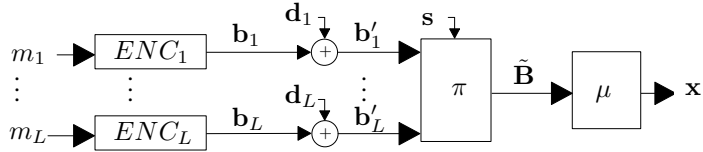}\\
  \caption{PBICM encoding scheme. `$+$' denoted modulo-2 addition (XOR). }\label{fig:BICMencoderDither}
\end{center}\end{figure}

Let the dither signals $\bd_l=[d_{l1},...,d_{ln}]$, $l \in \{1,...,L\}$ be $L$ random vectors, each of length $n$, that are drawn independently from a memoryless equiprobable binary source. The output of each encoder $ENC_l$, $\bb_l$, goes through a component-wise XOR operation with the dither vector $\bd_l$. The output of the XOR operation, denoted $\bb_l'$, is fed into the interleaver $\pi$. The full PBICM encoding scheme is shown in Fig. \ref{fig:BICMencoderDither}.

\begin{figure}\begin{center}
    \psfrag{&y}{$\by$}
    \psfrag{&llr}{$\! \! \! \! \! \! \! \begin{array}{c}
                        \textrm{\small LLR}\\
                        \textrm{\small calc.}
                  \end{array}$ }
    \psfrag{&Zt}{$\tilde{\bZ}$}
    \psfrag{&s}{$\bs$}
    \psfrag{&pi}{$\pi^{-1}$}
    \psfrag{&p}{$\!_{^\times}$}
    \psfrag{&z1}{$\bz_1$}
    \psfrag{&zL}{$\bz_L$}
    \psfrag{&z1p}{$\bz_1'$}
    \psfrag{&zLp}{$\bz_L'$}
    \psfrag{&d1}{$\!\!\boldsymbol\delta_1$}
    \psfrag{&dL}{$\!\!\boldsymbol\delta_L$}
    \psfrag{&D1}{$\! DEC_1$}
    \psfrag{&DL}{$\! DEC_L$}
    \psfrag{&vd}{$\vdots$}
    \psfrag{&m1}{$\hat m_1$}
    \psfrag{&mL}{$\hat m_L$}
    \psfrag{&x}{$\!\!\bx$}

  \includegraphics[width=4in]{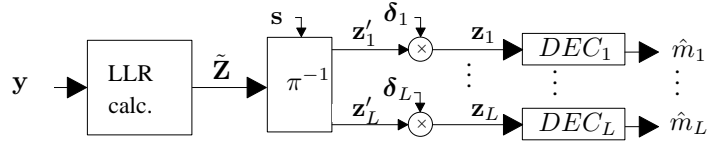}\\
  \caption{PBICM decoding scheme. $\boldsymbol\delta_l \triangleq 1-2\cdot \bd_l$, `$\times$' denotes element-wise multiplication.}\label{fig:BICMdecoderDither}
\end{center}\end{figure}

We let each decoder $DEC_l$ know the value of the dither used by its corresponding encoder $ENC_l$, $\bd_l$ (in practice the dither signals are generated using a pseudo-random generator which allows the common randomness). In order to compensate for the dither at the decoder, the LLR values are modified by flipping their sign for each dither value of 1 (and maintaining the sign where the dither is 0). Formally,  denote the LLR values at the de-interleaver output by $\bz_l'=[z_{l1}'\ ...\ z_{ln}']$. The LLR values at the decoders input shall be denoted by $\bz_l=[z_{l1}\ ...\ z_{ln}]$ and calculated as follows:
\begin{equation}
    z_{lj} =  z_{lj}' (1-2d_{lj}) ,\quad j=1,...,n.
\end{equation}
The PBICM decoding scheme is shown in Fig. \ref{fig:BICMdecoderDither}.

\subsection{Model equivalence}

Before we analyze the channel between each encoder-decoder pair in PBICM, let us define a binary memoryless channel that is related to $\Wbicm$, that will prove useful in the analysis of PBICM.

\begin{figure}\begin{center}
    \psfrag{&b}{$\quad \ B$}
    \psfrag{&S}{\!\!\!\!\!\!\!\!\!$S\in\!\{1,..L\}$}
    \psfrag{&s}{$S$}
    \psfrag{&D}{\!\!\!\!\!\!\!$D\in\!\{0,1\}$}
    \psfrag{&d}{$D$}
    \psfrag{&x}{$\ X$}
    \psfrag{&y}{$\!Y$}
    \psfrag{&ws}{$W_s$}
    \psfrag{&wb}{$\ \Wbicm$}
    \psfrag{&wbs}{$\ \ \WbicmSym$}

  \includegraphics[width=3.2in]{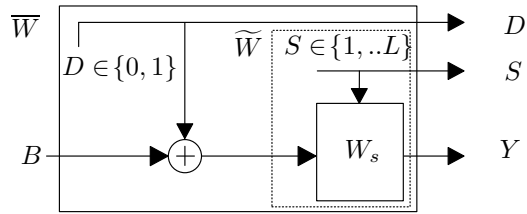}\\
  \caption{The binary channel $\WbicmSym$. The random state $S$ and the dither $D$ are known at the receiver.}\label{fig:WbicmSym}
\end{center}
\end{figure}

\begin{defn}\label{def:WbicmSym}
Let $\WbicmSym$ be a memoryless binary channel with input $B$ and output $\brkt{Y,S,D}$: $S$ is drawn at random from $\{1,...,L\}$, $D$ is drawn at random from $\{0,1\}$ ($S$ and $D$ are independent, and both do not depend on the input $B$). $Y$ is the output of the channel $W_S$ with input $B \oplus D$ ($\oplus$ is the XOR operation). Note that the channel $\WbicmSym$ is the channel $\Wbicm$ where the input is XORed with a binary RV $D$ (see Fig. \ref{fig:WbicmSym}).
\end{defn}
Note that the LLR calculation for the channel $\WbicmSym$ is given by
\begin{equation}\label{eqn:LLRWbicmSym}
    LLR_{\WbicmSym}(y,s,d) = (-1)^d LLR_{\Wbicm}(y,s) = (-1)^d LLR_s(y),
\end{equation}
where $LLR_{\Wbicm}$ and $LLR_{s}$ are given in (\ref{eqn:LLRWbicm}) and (\ref{eqn:LLRbicm}), respectively.

\begin{theorem}\label{thm:equivalence}
    In parallel BICM, the channel between every encoder-decoder pair is exactly the binary memoryless channel $\WbicmSym$, with its exact LLR output.
    \begin{proof}
    Consider the pair $ENC_1$ and $DEC_1$. Let $\bb_1$ be the codeword sent from $ENC_1$. After adding the dither $\bd_1$, the dithered codeword $\bb_1'$ enters the interleaver. The other codewords $\bb_2,...,\bb_L$ are dithered using $\bd_2,...,\bd_L$. Since the dither of these codewords is unknown at $DEC_1$, the dithered codewords $\bb_2',...,\bb_L'$ are truly random i.i.d. signals. The interleaving signal $\bs$ interleaves the dithered codewords according to (\ref{eqn:interleaver}). The interleaved signal enters the mapper $\mu$ and the channel $W$, resulting in an output $\by$. Since the dithered codewords $\bb_2',...,\bb_L'$ are i.i.d., the equivalent channel from $\bb_1'$ to $\brkt{\by,\bs}$ is exactly the channel $\Wbicm$. The LLR calculation at the PBICM receiver along with the interleaver produce  $\bz_1'$ , which is exactly the LLR calculation that fits the channel $\Wbicm$ (cf. (\ref{eqn:LLRWbicm})).

    Recalling that the channel $\WbicmSym$ is nothing but the channel $\Wbicm$ with its input XORed with a binary RV, and that the LLR of the channel $\Wbicm$ can be easily modified by the dither according to Eq. (\ref{eqn:LLRWbicmSym}) to produce the LLR of the channel $\WbicmSym$, we conclude that the channel between $\bb_1$ to $\bz_1$ is exactly the channel $\WbicmSym$ with $LLR$ calculation.

    Since by symmetry the above holds for any encoder-decoder pair $ENC_l$-$DEC_l$, the proof is concluded.
    \end{proof}
\end{theorem}

An important note should be made: Parallel BICM allows the decomposition of the nonbinary channel $W$ to $L$ binary channels of the type $\WbicmSym$. These $L$ channels are \emph{not} independent. For example, if $W$ is an additive noise channel, and at some point the noise instance is very strong, this will affect all the decoders and they will fail in decoding together. However, since in the PBICM scheme the channels are used independently, the operation of each decoder depends only on the marginal distribution of the relevant channel outputs. The outputs of these decoders will inevitably be statistically dependent, and we take this into consideration when analyzing the performance of coding using PBICM in the following.

\subsection{Error Probability Analysis}

We wish to analyze the performance of PBICM, and specifically, we are interested in the overall codeword error probability. Let $\C$ be a binary $(n,R)$ code, used in the PBICM scheme. To assure a fair comparison, we regard each $L$ consecutive information messages $(m_1,...,m_L)$ as a single message $m$, and regard the scheme as a code of length $n$ on the channel input alphabet $\X$. We define the following error events: Let $\Err_l$ be the event of a codeword error in $DEC_l$, and let $\Err$ be the event of an error in \emph{any} of the messages $\{m_1,...,m_L\}$, i.e. $\Err = \bigcup_l \Err_l$. Denote the corresponding error probabilities by $p_{e_l}$ and $p_e$ respectively.

\begin{cor}\label{cor:PBICMpe}
    Let $p_e(\WbicmSym)$ be the codeword error probability of the code $\C$ over the channel $\WbicmSym$. Then the overall error probability $p_e$ of the code $\C$ used with PBICM can be bounded by
    \begin{equation}\label{eqn:peEquivalenceBound}
        p_e(\WbicmSym) \leq p_e \leq L\cdot p_e(\WbicmSym).
    \end{equation}
    \begin{proof}
        Since the error events $\Err_l$ in codewords that are mapped to the same channel codeword together are dependent, we can only bound the overall error probability $p_e$ using the union bound. $p_e$ can be also lower bounded by the minimum of the error probabilities in any of the channels:
        \begin{equation}\label{eqn:peUBgeneral}
            \min \{p_{e_1},...,p_{e_L} \} \leq p_e \leq \sum_l p_{e_l}.
        \end{equation}

        Since by Theorem \ref{thm:equivalence} the channel between each of the encoder-decoder pairs is $\WbicmSym$, we get that the error probabilities must be all equal to the error probability of the code $\C$ over the channel $\WbicmSym$. Setting $p_{e_1}=p_{e_2}=...=p_{e_L}=p_e(\WbicmSym)$ in (\ref{eqn:peUBgeneral}) completes the proof.
    \end{proof}
\end{cor}

In many cases the bit error rate (BER) is of interest. Suppose that each of the messages $(m_1,...,m_L)$ represents $k$ information bits and the entire message $m$ represents $L\cdot k$ information bits. Let $\Err^b_{lk'}$ denote the error in the $k'$-th bit of the information message $m_l$. The average BER for the encoder-decoder pair $ENC_l$-$DEC_l$ is defined by

\begin{equation}
     p^b_{e_l} \triangleq \frac{1}{k} \sum_{k'=1}^{k} Pr\{\Err^b_{lk'}\}.
\end{equation}
Similarly, define the overall average BER as
\begin{equation}\label{eqn:AverageBER}
     p^b_e \triangleq \frac{1}{L\cdot k} \sum_{l=1}^{L} \sum_{k'=1}^{k} Pr\{\Err^b_{lk'}\} = \frac{1}{L}\sum_{l=1}^{L} p^b_{e_l}.
\end{equation}

\begin{cor}
    Let $p_e^b(\WbicmSym)$ be the average BER of a binary code $\C$ over the channel $\WbicmSym$. Then the average BER $p_e^b$ of the code $\C$ used with PBICM is equal to $p_e^b(\WbicmSym)$.
    \begin{proof}
        Follows directly from Theorem \ref{thm:equivalence} and from the definition of the average BER in (\ref{eqn:AverageBER}).
    \end{proof}
\end{cor}

\section{Parallel BICM: Information Theoretical Analysis}\label{sec:PBICM_IT}

In the previous section we defined the PBICM scheme and analyzed its basic error probability properties. The equivalence of the channel between each encoder-decoder pair that was established in Theorem \ref{thm:equivalence} enables a full information-theoretical analysis of the scheme. We show that the highest achievable rate by PBICM (the PBICM capacity) is equal to the BICM capacity as in Equation (\ref{eqn:BICMcapacity}), which should not be a surprise. At the finite-length regime, we derive error exponent and channel dispersion results as information-theoretical measures for optimal PBICM performance at finite-length.

\subsection{Capacity}

Let the PBICM capacity of $W$, $\bC^{\PBICM}(W)$,  be the highest achievable rate for reliable communication over the channel $W$ with PBICM and a given mapping $\mu$. (As usual, reliable communication means a vanishing codeword error probability as the codelength $n$ goes to infinity.)
\begin{theorem}\label{thm:PBICMcapacity}
    The PBICM capacity is given by
    \begin{equation}\label{eqn:PBICMcapacity}
        \bC^{\PBICM}(W) = L\cdot \bC(\WbicmSym) = \sum_{s=1}^L \bC(W_s) = \bC^{BICM}(W).
    \end{equation}
    \begin{proof}

        \emph{Achievability:} Let $\C^{(n)}$ be a series of (binary) capacity-achieving codes for the channel $\WbicmSym$, and let $p_e^{(n)}(\WbicmSym)$ be the corresponding (vanishing) codeword error probabilities. By Corollary \ref{cor:PBICMpe}, the overall error probability of PBICM with a binary code is upper bounded by $L$ times the error probability of the same code over the channel $\WbicmSym$, therefore when the codes $\C^{(n)}$ are used with PBICM, the overall error probability is bounded by $L\cdot p_e^{(n)}(\WbicmSym)$ and also vanish with $n$. Since there are $L$ instances of the channel $\WbicmSym$, we get that the rate of $L\cdot \bC(\WbicmSym)$ is achievable by PBICM.

        \emph{Converse:} Let $\C^{(n)}$ be a series of binary codes that are used with PBICM and achieve a vanishing overall error probability $p_e^{(n)}$, and suppose that the overall PBICM rate is given by $L\cdot R$ (a rate of $R$ at each encoder-decoder pair).  By Corollary \ref{cor:PBICMpe}, the codeword error probability of a code over $\WbicmSym$ is upper bounded by the overall error probability of the same code used in PBICM. Therefore, if $p_e^{(n)}$ vanishes as $n \ra \infty$, then the error probability over $\WbicmSym$ must also vanish, and therefore the communication rate between each encoder-decoder pair must be upper bounded by $\bC(\WbicmSym)$, and the overall rate cannot surpass $L\cdot \bC(\WbicmSym)$.

        All that remains is to calculate the capacity of $\WbicmSym$:
        \begin{equation}
            \bC(\WbicmSym) = I(B;Y,S,D) = I(B;Y,S|D) = \frac{1}{2} \left(I(B;Y,S|D=0)+ I(B;Y,S|D=1)\right).
        \end{equation}
        When $D=0$, we get the channel $\Wbicm$ exactly, and when $D=1$ we get the channel $\Wbicm$ with its input symbols always switched. In either way, the expression $I(B;Y,S|D=d)$ is equal to the capacity of $\Wbicm$. Using Lemma \ref{lem:CBICM}, we get that
        \begin{equation}
            \bC(\WbicmSym) = \bC(\Wbicm) = \frac{1}{L} \sum_{s=1}^L \bC(W_s).
        \end{equation}
    \end{proof}
\end{theorem}

A note regarding the capacity proof: one might me tempted to try and prove the capacity theorem for PBICM without dither, since with random coding, the code $\C$ is merely an i.i.d. binary random vector. This approach fails because of the following. In the decoding of each codeword, the correctness of the model $\Wbicm$ relies on the fact that the \emph{other} codewords are i.i.d. signals. Since PBICM requires a single code for all the $L$ levels, such a condition can never be met. It it possible to prove the achievability without dither when using a different random code at each level, but such an approach will not guarantee the existence of a single code, as required by PBICM.

\subsection{Error Exponent}
The error exponent of a channel $W$ is defined by
\begin{equation}\label{eqn:ErrExpDef}
    \E(R) \triangleq \lim_{n \ra \infty} -\frac{1}{n}\log\left(p_e(n)\right),
\end{equation}
where $p_e(n)$ is the average codeword error probability for the best code
of length $n$. A lower bound on the error exponent for memoryless channels is the \emph{random coding} error exponent \cite{GallagerInfoTheoryBook}, which is given by

\begin{equation}\label{eqn:Er}
    \E_r(R) = \max_{\rho \in [0,1]} \max_{P_X(\cdot)} \{ \E_0(\rho,P_X)-\rho R \},
\end{equation}
where
\begin{equation}\label{eqn:E0}
     \E_0(\rho,P_X) \triangleq -\log \left[ \sum_{y\in \Y} \left(\sum_{x\in\X} P_X(x)W(y|x)^{1/(1+\rho)}\right)^{1+\rho} \right].
\end{equation}
Since we consider equiprobable inputs only we omit the dependence of $\E_0(\rho)$ in $P_X$, and omit the maximization w.r.t. $P_X$ in (\ref{eqn:Er}).

Others known bounds on the error exponent include the
\emph{expurgation} error exponent lower bound, the \emph{sphere packing} error
exponent (an upper bound) and others \cite{GallagerInfoTheoryBook}. The random coding and sphere packing exponents coincide for rates above the critical rate, and therefore the error exponent is known precisely at these rates.

\subsubsection{PBICM error exponent} $ $

Similarly to (\ref{eqn:ErrExpDef}), we define the PBICM error exponent:
\begin{defn}
For a given channel $W$ and a mapping $\mu$, let $\E^{\PBICM}(R)$ be defined as
\begin{equation}\label{eqn:PBICMerrexpDef}
    \E^{\PBICM}(R) \triangleq \lim_{n \ra \infty} -\frac{1}{n}\log\left(p_e(n)\right),
\end{equation}
where $p_e(n)$ is the average codeword error probability for the best PBICM scheme with block length of $n$.
\end{defn}

Using Corollary \ref{cor:PBICMpe}, we can calculate the PBICM exponent using the error exponent of $\WbicmSym$:

\begin{theorem}\label{thm:PBICMerrexp}
The PBICM error exponent of a channel $W$ is given by
\begin{equation}\label{eqn:PBICMerrexp}
    \E^{\PBICM}(R) = \E(R/L),
\end{equation}
where $\E(\cdot)$ is the error exponent function of the binary channel $\WbicmSym$.

\begin{proof}Let $\C^{(n)}$ be a series of the binary codes. Denote their codeword error probabilities over the channel $\WbicmSym$ by $p_e^{(n)}(\WbicmSym)$. Let $p_e^{(n)}$ be the error probabilities of the corresponding PBICM schemes with $\C^{(n)}$ used as underlying codes. It follows from (\ref{eqn:peEquivalenceBound}) that
\begin{equation}
  -\frac{1}{n} \log (L\cdot  p_e^{(n)}(\WbicmSym))\leq  -\frac{1}{n} \log p_e^{(n)} \leq -\frac{1}{n} \log p_e^{(n)}(\WbicmSym).
\end{equation}
By taking $n \ra \infty$ the factor of $L$ vanishes and we get that for any series of codes,
\begin{equation}
    \lim _{n \ra \infty} -\frac{1}{n} \log p_e^{(n)} =\lim _{n \ra \infty} -\frac{1}{n} \log p_e^{(n)}(\WbicmSym).
\end{equation}
The above equation holds for the series of best codes for the channel $\WbicmSym$, as well as for the series of the best codes for PBICM. Therefore the equality holds for the sequence of best codes on either side. Since the rate for PBICM is $L$ times the rate for coding on $\WbicmSym$, the proof is concluded.
\end{proof}
\end{theorem}

\subsubsection{The error exponent of $\WbicmSym$} $ $

The channel $\WbicmSym$ has a special structure, and is related to the binary sub-channels $W_i$. We now calculate two basic bounds for the error exponent of $\WbicmSym$ in terms of the sub-channels $W_i$. By Theorem \ref{thm:PBICMerrexp}, the PBICM error exponent of the channel $W$ can be bounded accordingly.

\begin{theorem}\label{thm:WbicmSymExp}
Let $\E(R)$ be the error exponent of the channel $\WbicmSym$. It can be bounded as follows:

\emph{Random coding:}
\begin{equation}\label{eqn:WbicmSymExp}
    \E(R) \geq \E_r(R) = \max_{\rho \in [0,1]} \{ \E_0(\rho)-\rho R \},
\end{equation}
where
\begin{equation}\label{eqn:E0connection}
    \E_0(\rho) = -\log \EE \left[2^{- \E^{(\!S)}_0(\rho)}\right],
\end{equation}
$\E_0^{(\!s)}(\rho)$ is the $\E_0$ function for the channel $W_s$, and the expectation is w.r.t. the state $S$ which is drawn uniformly from $\{1..L\}$.

\emph{Sphere packing:}
\begin{equation}\label{eqn:WbicmSymExp}
    \E(R) \leq \E_{sp}(R) = \max_{\rho > 0} \{ \E_0(\rho)-\rho R \},
\end{equation}
where $\E_0(\rho)$ is given in (\ref{eqn:E0connection}).

\begin{proof}

The bounds in the theorem are the original random coding and sphere packing exponents \cite{GallagerInfoTheoryBook}. The proof, therefore, boils down to the simplification of the $\E_0$ function to the form of (\ref{eqn:E0connection}).

Consider the channel $\WbicmSym$ (Definition \ref{def:WbicmSym}) with binary input $B$ and outputs $\brkt{Y,S,D}$. Since $\WbicmSym$ is equivalent to the channel $\Wbicm$ with input $B\oplus D$, where $D$ is an equiprobable binary RV (and known at the receiver), we get that
\begin{equation}
    \WbicmSym(y,s,d|b) = \frac{1}{2}\Wbicm(y,s|b\oplus d).
\end{equation}
The channel $\Wbicm$, in turn, is nothing more than the channel $W_s$ with the additional output $S$. This yields
 \begin{equation}
    \frac{1}{2}\Wbicm(y,s|b\oplus d) = \frac{1}{2L}W_s(y|b\oplus d).
 \end{equation}

Combining the above, the function $\E_0$ of $\WbicmSym$ is therefore given by
\begin{eqnarray}\label{eqn:E0new}
    \E_0(\rho) & =&  -\log  \sum_{\overset{y\in \Y}{
    \underset{d \in \{0,1\}}{s \in  \{1..L\} }
    }} \nhs\left[\sum_{b\in\{0,1\}} P_B(b)\WbicmSym(y,s,d|b)^\frac{1}{1+\rho}\right]^{1+\rho} \nonumber \\
    & =&  -\log  \sum_{\overset{y\in \Y}{
    \underset{d \in \{0,1\}}{s \in  \{1..L\} }
    }} \nhs\left[\sum_{b\in\{0,1\}} \frac{1}{2}\left(\frac{1}{2L}W_s(y|b\oplus d)\right)^\frac{1}{1+\rho}\right]^{1+\rho} \nonumber \\
    &\overset{(a)}{=}&  -\log
    \sum_{\overset{y\in \Y}{s \in  \{1..L\}}} \nhs\left[\sum_{b'\in\{0,1\}}\frac{1}{2} \left(\frac{1}{L}W_s(y|b')\right)^\frac{1}{1+\rho}\right]^{1+\rho} \label{eqn:WbicmE0def} \\
    &=&  -\log
    \sum_{s \in  \{1..L\}} \frac{1}{L}
    \sum_{y\in \Y} \nhs\left[\sum_{b'\in\{0,1\}}\frac{1}{2} W_s(y|b')^\frac{1}{1+\rho}\right]^{1+\rho} \nonumber \\
    &\overset{(b)}{=}&  -\log
    \sum_{s \in  \{1..L\}} \frac{1}{L}
    2^{- \E_0^{(\!s)}(\rho)}\nonumber\\
    &=&  -\log \EE \left[ 2^{- \E_0^{(\!s)}(\rho)}\right]
\end{eqnarray}
$(a)$ follows by setting $b' \triangleq b\oplus d$, and by noting that the summation result is independent of the value of $d$. $(b)$ follows from the definition of $\E_0^{(\!s)}(\rho)$ (the $\E_0$ function for the channel $W_s$).

\end{proof}
\end{theorem}

Several notes can be made:
\begin{itemize}
    \item It is well known that the random coding and sphere packing exponents coincide at rates above the critical rate. Therefore the exact error exponent of $\WbicmSym$ is known at rates above the critical rate of $\WbicmSym$, $R_{cr}^{\WbicmSym}$. It follows that the exact PBICM error exponent is known at rates above $R_{cr}^{\PBICM} \triangleq L \cdot R_{cr}^{\WbicmSym}$, which we define to be the PBICM critical rate.
    \item In theorem \ref{thm:WbicmSymExp} we have shown that the random coding and the sphere packing bounds have a compact form because of the special structure of the channel $\WbicmSym$. Clearly, following Theorem \ref{thm:PBICMerrexp}, every bound on $\E(R)$ of $\WbicmSym$ serves as a bound on the PBICM error exponent. However, for other bounds (such as the expurgation error exponent \cite{GallagerInfoTheoryBook}), no compact form could be found. Such bounds, of course, can still be applied to bound $\E^{\PBICM}(R)$.
    \item The $\E_0$ function of the channel $\Wbicm$ is equal to the $\E_0$ function of the channel $\WbicmSym$. This can easily be seen from the proof above: $\E_0$ for $\Wbicm$ is given in (\ref{eqn:WbicmE0def}) by definition.
    \item In \cite{WachsmannFischerHuber_MLC_99}, the authors offered the model of $\Wbicm$ for calculating the error exponent of BICM. It is claimed that $\E_0$ of the channel $\Wbicm$ is given by \cite[Eq. (37)]{WachsmannFischerHuber_MLC_99}:
        \begin{equation}
            \EE\left[\E_0^{(\!S)}(\rho)\right] = \frac{1}{L}\sum_{s=1}^L \E_0^{(\!s)}(\rho).
        \end{equation}
        As we have just shown in Theorem \ref{thm:WbicmSymExp}, this is not the exact expression. In fact, it can be shown that $\E_0(\rho) \leq \EE\left[\E_0^{(\!S)}(\rho)\right]$. This follows directly from the convexity of the function $2^{-(\cdot)}$ and the Jensen inequality. Therefore the incorrect expression in \cite[Eq. (37)]{WachsmannFischerHuber_MLC_99} always overestimates the value of $\E_0(\rho)$, and therefore the resulting $\E_r(R)$ expression also overestimates the true random coding expression.
\end{itemize}

\subsection{Channel Dispersion}
An alternative information theoretical measure for quantifying coding performance with finite block lengths is the \emph{channel dispersion}.
Suppose that a fixed codeword error probability $p_e$ and a codeword length $n$ are given. We can then seek the maximal achievable rate $R$ given $p_e$ and $n$.

It appears that for fixed $p_e$ and $n$, the gap to the channel capacity is approximately proportional to $Q^{-1}(p_e)/\sqrt{n}$ (where $Q(\cdot)$ is the complementary Gaussian cumulative distribution function).
 The proportion constant (squared) is called the channel dispersion. Formally, define the (operational) channel dispersion as follows \cite{PolyanskiyPVFiniteLength10}:

\begin{defn}\label{def:dispersion}
The dispersion $\bV(W)$ of a channel $W$ with capacity $C$ is defined as
\begin{equation}\label{eqn:defDispersion}
    \bV(W) = \lim_{p_e \ra 0} \limsup_{n \ra \infty} \ n\cdot \left(\frac{C - R(n,p_e)}{Q^{-1}(p_e)}\right)^2,
\end{equation}
where $R(n,p_e)$ is the highest achievable rate for codeword error probability $p_e$ and codeword length $n$.
\end{defn}

In 1962 , Strassen \cite{Strassen62_Asymptotische} used the Gaussian approximation to derive the following result for DMCs\footnote{see Appendix \ref{app:notation} for the big-O notation.}:
\begin{equation}\label{eqn:dispersion}
    R(n,p_e) = C - \sqrt{V/n}Q^{-1}(p_e) + O\left(\frac{\log n}{n}\right),
\end{equation}
where $C$ is the channel capacity, and the new quantity $V$ is the (information-theoretic) dispersion , which is given by
\begin{equation}\label{eqn:dispersion_info_def}
    V \triangleq  \VAR (i(X;Y)),
\end{equation}
where $i(x;y)$ is the information spectrum, given by
\begin{equation}
     i(x;y) \triangleq  \log \frac{P_{XY}(x,y)}{P_X(x) P_Y(y)},
\end{equation}
and the distribution of $X$ is the capacity-achieving distribution that minimizes $V$.
Strassen's result proves that the dispersion of DMCs is equal to $\VAR(i(X;Y))$. This result was recently tightened (and extended to the power-constrained AWGN channel) in \cite{PolyanskiyPVFiniteLength10}. It is also known that the channel dispersion and the error exponent are related as follows. For a channel with capacity $C$ and dispersion $V$, the error exponent can be approximated by $\E(R) \cong \frac{(C-R)^2}{2V\ln 2}$. See \cite{PolyanskiyPVFiniteLength10} for details on the early origins of this approximation by Shannon.

\subsubsection{PBICM dispersion} $ $
In order to estimate the finite-block performance of PBICM schemes we extend the dispersion definition as follows:
\begin{defn}\label{def:PBICMdispersion}
The PBICM dispersion $\bV^{\PBICM}(W)$ of a channel $W$ with a given mapping $\mu$ and PBICM capacity $\bC^{\PBICM}(W)$ is defined as
\begin{equation}\label{eqn:defPBICMDispersion}
    \bV^{\PBICM}(W) = \lim_{p_e \ra 0} \limsup_{n \ra \infty} \ n\cdot \left(\frac{\bC^{\PBICM}(W) - R(n,p_e)}{Q^{-1}(p_e)}\right)^2,
\end{equation}
where $R(n,p_e)$ is the highest achievable rate for any PBICM scheme with a given $n$ and $p_e$.
\end{defn}

Relying on the relationship between the PBICM scheme and the binary channel $\WbicmSym$, we can show the following:

\begin{theorem}\label{thm:PBICMdispersion}
Let $n$ be a given block length and let $p_e$ be a given codeword error probability. The highest achievable rate attained using PBICM, $R^{\PBICM}(n,p_e)$ is bounded from above and below by:
\begin{eqnarray}
    R^{\PBICM}(n,p_e) &\geq&  \bC^{\PBICM}(W) - \sqrt{\frac{L^2\bV(\WbicmSym)}{n}}Q^{-1}\left(\frac{p_e}{L}\right) + O\left(\frac{1}{n}\right),\label{eqn:PBICMdispersionLower}\\
    R^{\PBICM}(n,p_e) &\leq& \bC^{\PBICM}(W) - \sqrt{\frac{L^2\bV(\WbicmSym)}{n}}Q^{-1}(p_e)+ O\left(\frac{\log n}{n}\right).\label{eqn:PBICMdispersionUpper}
\end{eqnarray}
As a result, the PBICM dispersion is given by
\begin{equation}\label{eqn:PBICMdispersionTheorem}
    \bV^{\PBICM}(W) = L^2 \bV(\WbicmSym).
\end{equation}
\begin{proof}

\emph{Direct:} From the achievability proof of (\ref{eqn:dispersion}) \cite[Theorem 45]{PolyanskiyPVFiniteLength10}, there must exist an $(n,R',p_e' = p_e/L)$ binary code for $\WbicmSym$ that satisfies
\begin{equation}
    R' \geq \bC(\WbicmSym) - \sqrt{\frac{\bV(\WbicmSym)}{n}}Q^{-1}(p_e/L) + O\left(\frac{1}{n}\right).
\end{equation}
By Theorem \ref{thm:equivalence} and Corollary \ref{cor:PBICMpe}, it follows that the PBICM scheme based on this code is not greater than  $L p_e' = p_e$. The rate of the PBICM scheme satisfies
\begin{eqnarray}
  R &=& L\cdot R' \geq L\left[\bC(\WbicmSym) - \sqrt{\frac{\bV(\WbicmSym)}{n}}Q^{-1}(p_e') + O\left(\frac{1}{n}\right)\right] \\
    &=& \bC^{BICM}(W) - \sqrt{\frac{L^2 \bV(\WbicmSym)}{n}}Q^{-1}\left(\frac{p_e}{L}\right) + O\left(\frac{1}{n}\right).
\end{eqnarray}

\emph{Converse:} Suppose we have a $(n,R,p_e)$ PBICM scheme. According to Corollary \ref{cor:PBICMpe}, the codeword error probability $p_e'$ of the underlying binary code is not greater than than $p_e$. By Equation (\ref{eqn:dispersion}), the rate $R'$ of the underlying binary code is bounded by
\begin{equation}
    R' \leq \bC(\WbicmSym) - \sqrt{\frac{\bV(\WbicmSym)}{n}}Q^{-1}(p_e') + O\left(\frac{\log n}{n}\right).
\end{equation}
Since $Q^{-1}(\cdot)$ is a decreasing function, the bound loosens by replacing $p_e'$ with the higher $p_e$. Therefore the overall rate $R$ is bounded by
\begin{eqnarray}
  R &=& L\cdot R' \leq L\left[\bC(\WbicmSym) - \sqrt{\frac{\bV(\WbicmSym)}{n}}Q^{-1}(p_e') + O\left(\frac{\log n}{n}\right) \right]\\
    &\leq& \bC^{\PBICM}(W) - \sqrt{\frac{L^2\bV(\WbicmSym)}{n}}Q^{-1}\left(p_e\right)+O\left(\frac{\log n}{n}\right).
\end{eqnarray}

\emph{PBICM dispersion:} Rewriting Equations (\ref{eqn:PBICMdispersionLower}) and (\ref{eqn:PBICMdispersionUpper}), we get the following:

\begin{equation}
 \sqrt{\frac{L^2\bV(\WbicmSym)}{n}}Q^{-1}(p_e)+ O\left(\frac{\log n}{n}\right)  \leq  \bC^{\PBICM}(W) -R \leq  \sqrt{\frac{L^2\bV(\WbicmSym)}{n}}Q^{-1}\left(\frac{p_e}{L}\right) +O\left(\frac{1}{n}\right)
\end{equation}

\begin{equation}
 \sqrt{L^2\bV(\WbicmSym)}+ O\left(\frac{\log n}{\sqrt n}\right)  \leq  \sqrt{n} \left(\frac{\bC^{\PBICM}(W) -R}{Q^{-1}(p_e)}\right) \leq \sqrt{L^2\bV(\WbicmSym)} \cdot \frac{Q^{-1}\left(\frac{p_e}{L}\right)}{Q^{-1}(p_e)} +O\left(\frac{1}{\sqrt n}\right)
\end{equation}

Taking the limit w.r.t. $n$ yields

\begin{equation}
 \sqrt{L^2\bV(\WbicmSym)} \leq  \limsup_{n \ra \infty}\sqrt{n} \left(\frac{\bC^{\PBICM}(W) -R}{Q^{-1}(p_e)}\right) \leq \sqrt{L^2\bV(\WbicmSym)} \cdot \frac{Q^{-1}\left(\frac{p_e}{L}\right)}{Q^{-1}(p_e)},
\end{equation}
or
\begin{equation}
 L^2\bV(\WbicmSym) \leq  \limsup_{n \ra \infty}n \left(\frac{\bC^{\PBICM}(W) -R}{Q^{-1}(p_e)}\right)^2 \leq L^2\bV(\WbicmSym) \left(\frac{Q^{-1}\left(\frac{p_e}{L}\right)}{Q^{-1}(p_e)}\right)^2.
\end{equation}

By noting that $\lim_{\eps \ra 0^+} \frac{Q^{-1}(\eps)^2}{2\ln\frac{1}{\eps}} = 1$ (see Appendix \ref{app:invQapprox}),
we get that
\begin{equation}
    \lim_{p_e \ra 0} \left(\frac{Q^{-1}\left(\frac{p_e}{L}\right)}{Q^{-1}(p_e)}\right)^2=\lim_{p_e \ra 0} \frac{\ln(L/p_e)}{\ln(1/p_e)}= \lim_{p_e \ra 0} \frac{\ln(1/p_e)+\ln L}{\ln(1/p_e)} = 1,
\end{equation}
which leads to the desired result:
\begin{equation}
    \bV^{\PBICM}(W) = \lim_{p_e \ra 0} \limsup_{n \ra \infty} \ n\cdot \left(\frac{\bC^{\PBICM}(W) - R(n,p_e)}{Q^{-1}(p_e)}\right)^2 = L^2\bV(\WbicmSym).
\end{equation}

\end{proof}
\end{theorem}

Note that the PBICM dispersion result is not as tight as the bound for general coding schemes as in (\ref{eqn:dispersion}). The reason is the unavoidable use of the union bound when estimating the overall error probability of PBICM in Theorem \ref{thm:equivalence}. In the dispersion proof for DMCs, the value of the dispersion is obtained even without taking the limit w.r.t. $p_e$. However, the gap between $Q^{-1}(p_e)$ and $Q^{-1}(p_e/L)$ for values of interest is not very large.

\subsubsection{The dispersion of $\WbicmSym$} $ $

As in the error exponent case, the PBICM dispersion of a channel is related to the dispersion of the binary channel $\WbicmSym$. We now calculate it explicitly from the dispersions of the sub-channels $W_i$.

\begin{theorem}\label{thm:WbicmDisp}
The dispersion of the channel $\WbicmSym$ is given by
\begin{equation}
    \bV(\WbicmSym) = \bV(\Wbicm)= 
    \EE [ \bV(W_S)] + \VAR\left[ \bC(W_S)\right] = \left[\frac{1}{L}\sum_{s=1}^L \bV(W_s)\right] + \VAR(\bC(W_S))
\end{equation}
where $\VAR(\bC(W_S))$ is the statistical variance of the capacity of $W_s$, i.e.
\begin{equation}
    \VAR(\bC(W_S)) \triangleq \EE[\bC^2(W_S)] - \EE^2[\bC(W_S)].
\end{equation}
\begin{proof}
Consider the channel $\WbicmSym$ (Definition \ref{def:WbicmSym}) with binary input $B$ and outputs $\brkt{Y,S,D}$, and recall that
\begin{equation}
    P_{YSD|B}(y,s,d|b) = \WbicmSym(y,s,d|b) = \frac{1}{2}\Wbicm(y,s|b\oplus d) = \frac{1}{2L}W_s(y|b\oplus d).
\end{equation}

We first calculate the dispersion of $\Wbicm$. Since $S$ and the channel input $B$ are independent, the information spectrum is given by

\begin{eqnarray}\label{eqn:info_spec_SI}
    i(b;y,s) &\triangleq& \log \frac{P_{YSB}(y,s,b)}{P_{YS}(y,s)P_B(b)} = \log \frac{P_{Y|SB}(y|s,b)P_S(s)P_B(b)}{P_{YS}(y,s)P_B(b)} \\
    &=&\log \frac{P_{Y|SB}(y|s,b)}{P_{Y|S}(y|s)} \triangleq i(b;y|s).
\end{eqnarray}

Using this notation, the dispersion of the channel $W_s$ is given by
\begin{eqnarray*}
   \bV(W_s)\hspace{-.1in} &=&\hspace{-.1in} \VAR( i(B;Y|s) | S=s)\\
    &=& \EE\left[i^2(B;Y|s) | S=s\right] -\bC(W_s)^2.
\end{eqnarray*}

Next, the dispersion of the channel $\Wbicm$ is given as follows:
\begin{eqnarray}
    \bV(\Wbicm) \hspace{-.1in} &=& \VAR( i(B;Y,S)) = \VAR(i(B;Y|S))\nonumber\\
    &\overset{(a)}{=}& \EE \left[ \VAR [i(B;Y|s) | S=s]\right]  + \VAR\left[ \EE[i(B;Y|S) | S=s]\right]\nonumber\\
    &=& \EE [ \bV(W_S)] + \VAR\left[ \bC(W_S)\right] = \left[\frac{1}{L}\sum_{s=1}^L \bV(W_s)\right] + \VAR(\bC(W_S)).\label{eqn:dispersion_SI}
\end{eqnarray}
$(a)$ follows from the law of total variance.

Finally, the dispersion of the channel $\WbicmSym$ is calculated as follows:

Let us combine the outputs of the channel $\Wbicm$ to a single output $Z=\brkt{Y,S}$. We therefore end up with a channel with input $B$ and outputs $Z$ and $D$ (see Fig. \ref{fig:WbicmSym}). Similarly to (\ref{eqn:info_spec_SI}), we get that the information spectrum is given by
\begin{equation}
        i(b;z,d) \triangleq \log \frac{P_{ZDB}(z,d,b)}{P_{ZD}(z,d)P_B(b)} = i(b;z|d).
\end{equation}
Following (\ref{eqn:dispersion_SI}), we get that
\begin{equation}
    \bV(\WbicmSym) = \EE [ \bV(\Wbicm_D)] + \VAR\left[ \bC(\Wbicm_D)\right] = \frac{1}{2}\sum_{d=\{0,1\}} \bV(\Wbicm_d) + \VAR(\bC(\Wbicm_D)),
\end{equation}
where $\Wbicm_d$ is the channel $\Wbicm$ with its input XORed with the value $d$.

Since only equiprobable inputs are considered, it follows that $\bC(\Wbicm_0) = \bC(\Wbicm_1) = \bC(\Wbicm)$, and that $\bV(\Wbicm_0) = \bV(\Wbicm_1) = \bV(\Wbicm)$. It therefore follows that $\VAR(\bC(\Wbicm_D)) = 0$, and consequently, $\bV(\WbicmSym) = \bV(\Wbicm)$, as required.
\end{proof}
\end{theorem}

Note that since large dispersion means higher backoff from the capacity (see (\ref{eqn:dispersion})), the term $\VAR(\bC(W_S))$ can be thought of as a \emph{penalty factor} for the dispersion, over the expected dispersion over the channels $W_s$, $\EE[\bV(W_S)]$. This factor grows as the capacities of the sub-channels $W_i$ are more spread.

\section{Numerical Results}\label{sec:numerics}

In this section we evaluate numerically the information-theoretical quantities for PBICM. In particular, we calculate the PBICM random coding error exponent (see Theorems \ref{thm:PBICMerrexp} and \ref{thm:WbicmSymExp}) in order to compare with the mismatched decoding approach \cite{MartinezFabregas09_BICM_mismatched}. We consider the AWGN channel and the Rayleigh fading channel (with perfect channel state information at the receiver) over a wide range of SNR values and constellations. Gray mapping was used throughout all the examples.

\subsection{Normalization: latency vs. complexity}
One way to compare the PBICM error exponent with the mismatched decoding exponent is to compare the error probability when the block length $n$ is fixed, which gives a simple comparison between the exponent values. Such an approach makes sense, since both schemes have the same latency of $n$ channel uses. As will be seen in the coming examples, for fixed $n$ the PBICM error exponent is inferior to that of the mismatched decoding. This can also be seen by observing that the PBICM random coding exponent has a slope of $-1/L$ (in its straight-line region), where the mismatched decoding exponent has a slope of $-1$.

However, it should be taken into consideration that when the block length is $n$, the mismatched decoder is working with a binary code of length $n\cdot L$. The complexity of the maximum-metric decoder is proportional to the number of codewords $2^{n\cdot L\cdot R}$ \cite{MartinezFabregas09_BICM_mismatched}, where $R$ is the rate of the binary code. On the other hand, the number of codewords in the PBICM scheme is $L \cdot 2^{n\cdot R}$ only. In order to assure a fair comparison from the complexity point of view, one has to allow the PBICM scheme to use a block length that is $L$ time the block length of the mismatched decoding scheme. Comparing the error probabilities of both schemes gives $nL\E_r^{\textrm{\PBICM}} =n\E_r^{\textrm{Mismatched}}$. We therefore define the normalized PBICM error exponent as $L$ times the PBICM error exponent. We conclude that when the complexity is more important (and the latency is less important), the normalized PBICM exponent is the quantity of interest.

It could be claimed, of course, that practical codes used today (such as low-density parity check (LDPC) codes ) will be used and they do not have exponential decoding complexity. On the other hand, such codes do not guarantee an exponentially decaying error probability.

\subsection{Comparison with the Mismatched Decoding Exponent}

In the following figures we show the comparison between the PBICM error exponent and the mismatched decoding error exponent \cite{MartinezFabregas09_BICM_mismatched}. The figures show the (unconstrained) random coding error exponent of the channel, along with the mismatched error exponent and the PBICM random coding error exponent (both normalized and un-normalized).

\begin{figure}\begin{center}
  \includegraphics[width=5in]{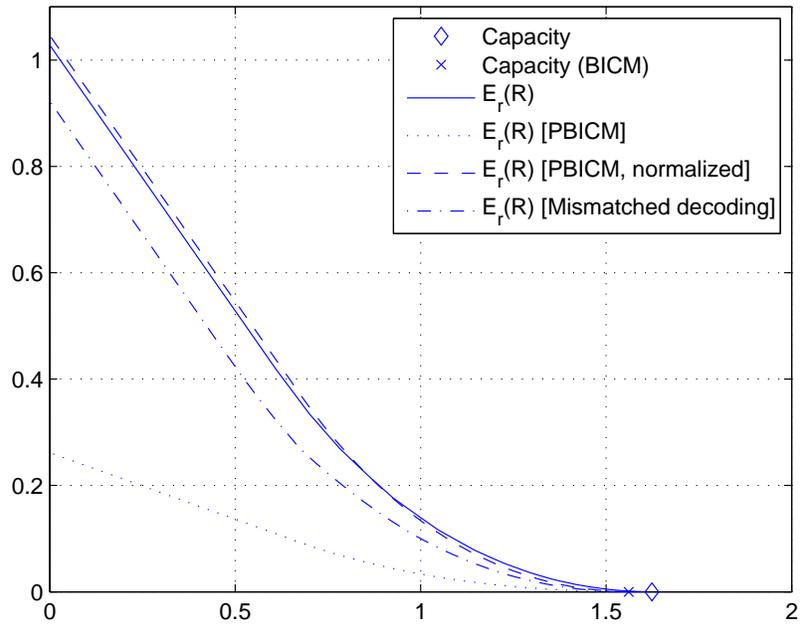}\\
  \caption{Random coding exponents over the Rayleigh fading channel with 16-QAM signaling and SNR of 5dB. }\label{fig:Rayleigh5dB16QAM}
\end{center}\end{figure}

\begin{figure}\begin{center}
  \includegraphics[width=5in]{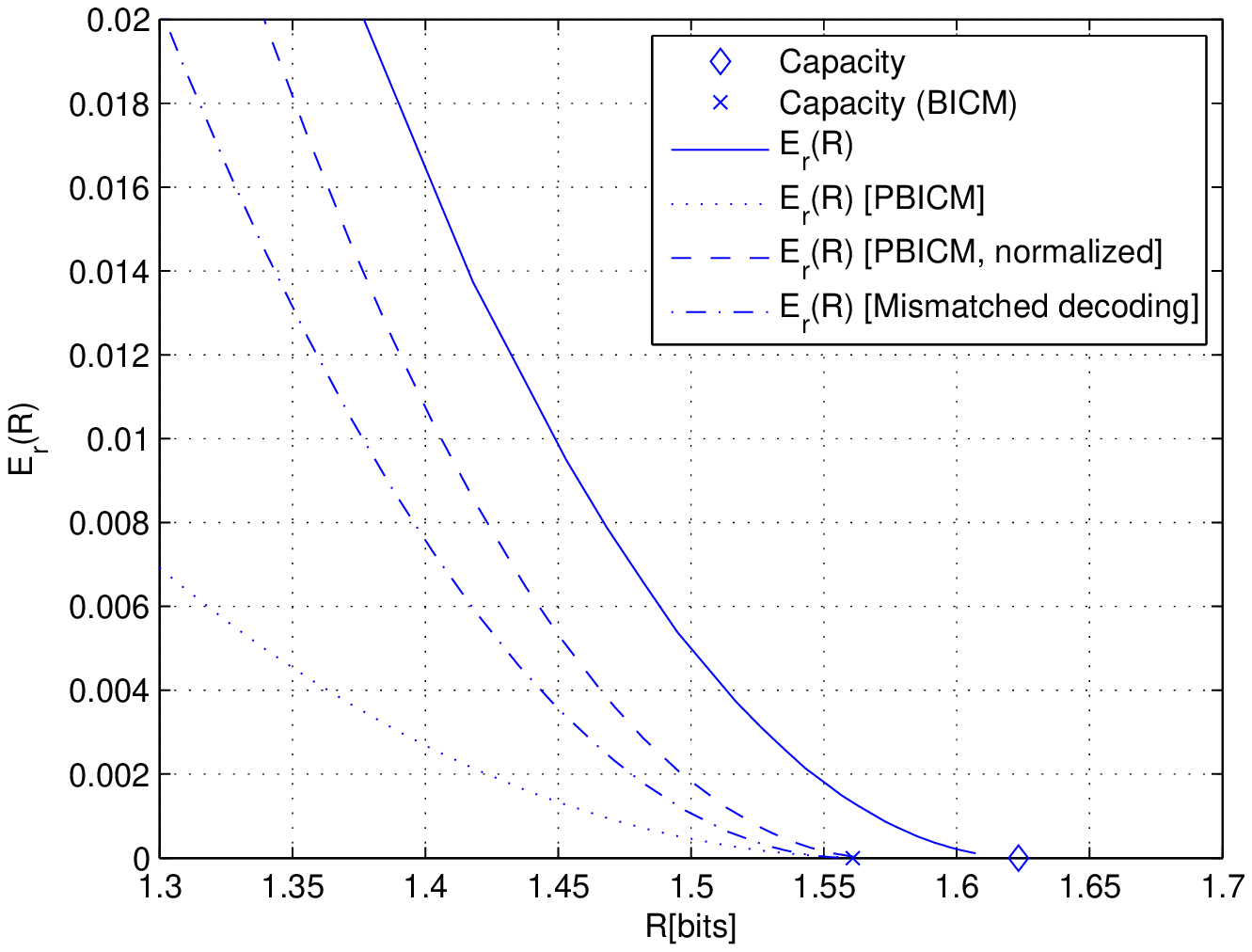}\\
  \caption{Random coding exponents over the Rayleigh fading channel with 16-QAM signaling and SNR of 5dB (zoomed on the capacity) }\label{fig:Rayleigh5dB16QAMzoom}
\end{center}\end{figure}

Figure \ref{fig:Rayleigh5dB16QAM} compares the exponents of 16QAM signaling over the Rayleigh fading channel at SNR = 5dB. Figure \ref{fig:Rayleigh5dB16QAMzoom} shows the same graph, zoomed-in on the capacity region. It can be seen that throughout the entire range of rates between zero and the BICM capacity, the normalized PBICM random coding exponent is higher (better) than the mismatched decoding exponent. Both BICM exponents are above zero for rates below the BICM capacity, and the unconstrained random coding exponent reaches zero at the full channel capacity, as expected. A fact that might be somewhat surprising at first glance is that the normalized PBICM exponent is better than the unconstrained random coding exponent in some rates. While this may seem contradictory, recall that we consider coding schemes with the same maximum-likelihood (or maximum metric) complexity. When normalizing the schemes complexity, PBICM operates with a block length that is $L$ times the block length of the unconstrained scheme, and therefore there is no contradiction. The mismatched decoder never attains higher values than the unconstrained exponent, a fact that is known as the data processing inequality for exponents (see e.g. \cite[Proposition 3.2]{MartinezFabregas09_BICM_mismatched}).

\begin{figure}\begin{center}
  \includegraphics[width=5in]{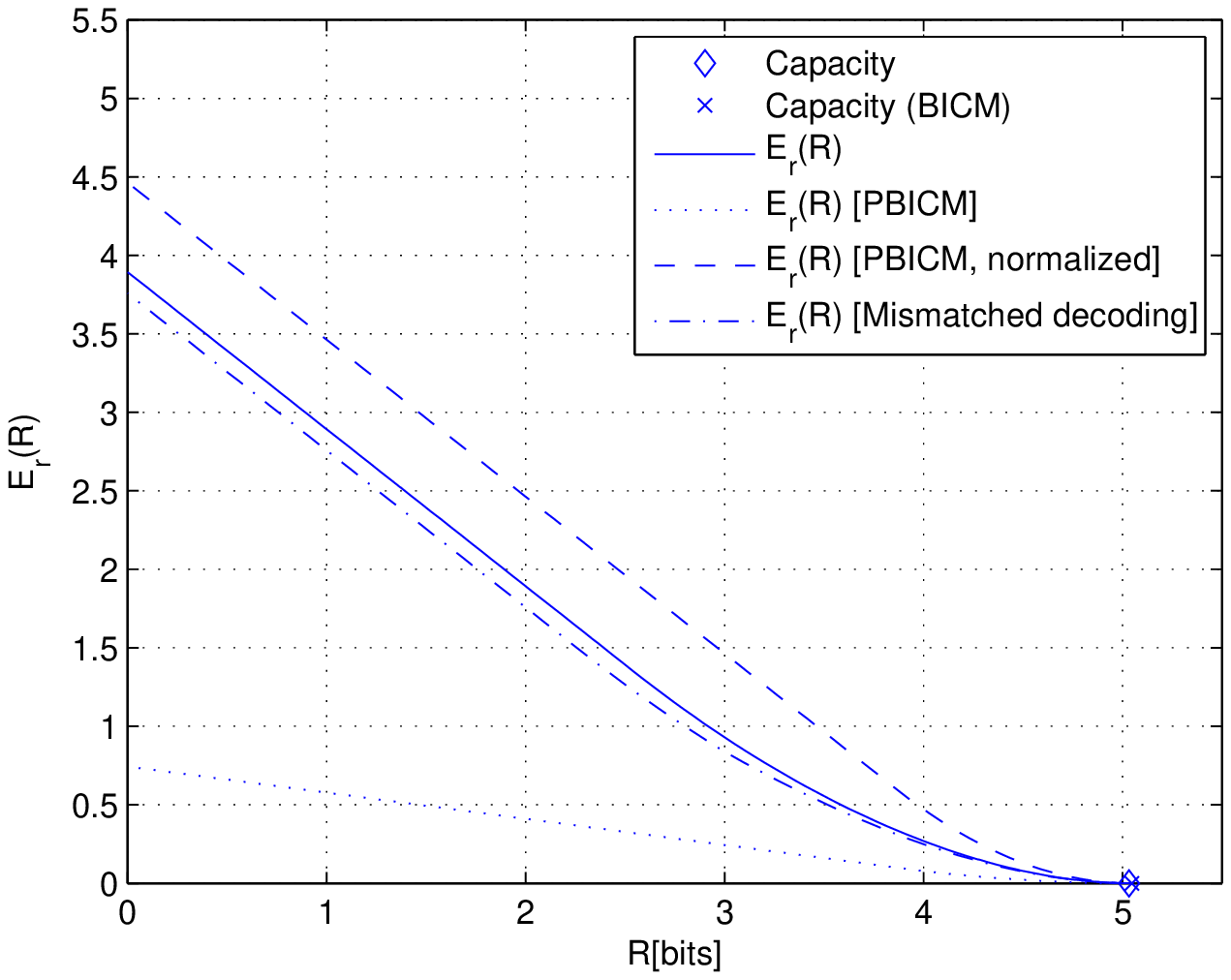}\\
  \caption{Random coding exponents over the Rayleigh fading channel with 64-QAM signaling and SNR of 20dB. }\label{fig:Rayleigh20dB64QAM}
\end{center}\end{figure}

\begin{figure}\begin{center}
  \includegraphics[width=5in]{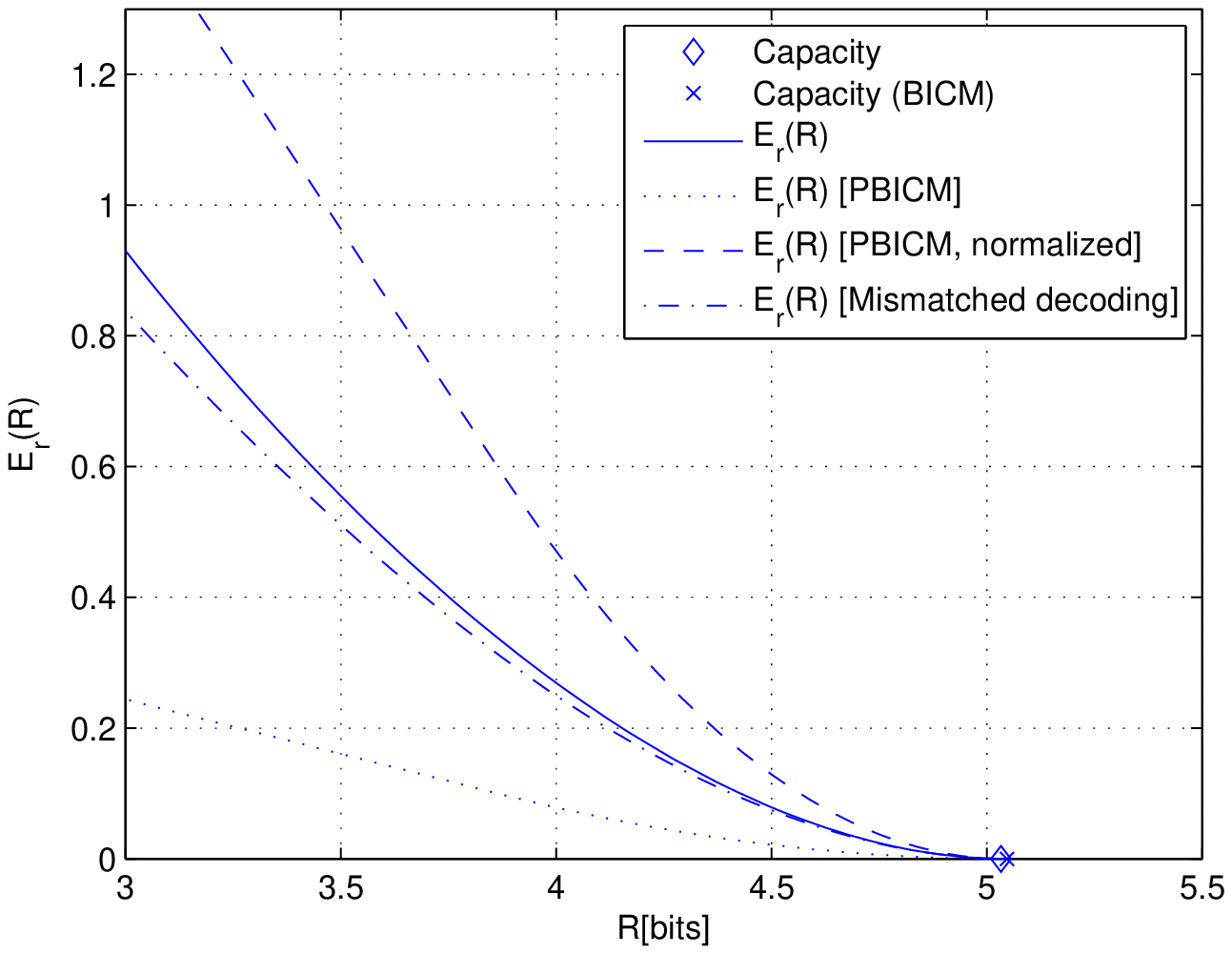}\\
  \caption{Random coding exponents over the Rayleigh fading channel with 64-QAM signaling and SNR of 20dB (zoomed on the capacity) }\label{fig:Rayleigh20dB64QAMzoom}
\end{center}\end{figure}

Figure \ref{fig:Rayleigh20dB64QAM} shows a similar picture (zoomed on the capacity in Figure \ref{fig:Rayleigh20dB64QAMzoom}). Again, the normalized PBICM outperforms the mismatched decoding exponent for all rates. In this case, the BICM capacity is very close to the full channel capacity, which enables the normalized PBICM to outperform the unconstrained exponent for essentially all rates.

On the Rayleigh fading channel, the same behavior was observed for the range of all practical ranges of SNR for 8PSK, 16QAM and 64QAM signaling: the normalized PBICM exponent outperformed the mismatched decoding exponent.

\begin{figure}\begin{center}
  \includegraphics[width=5in]{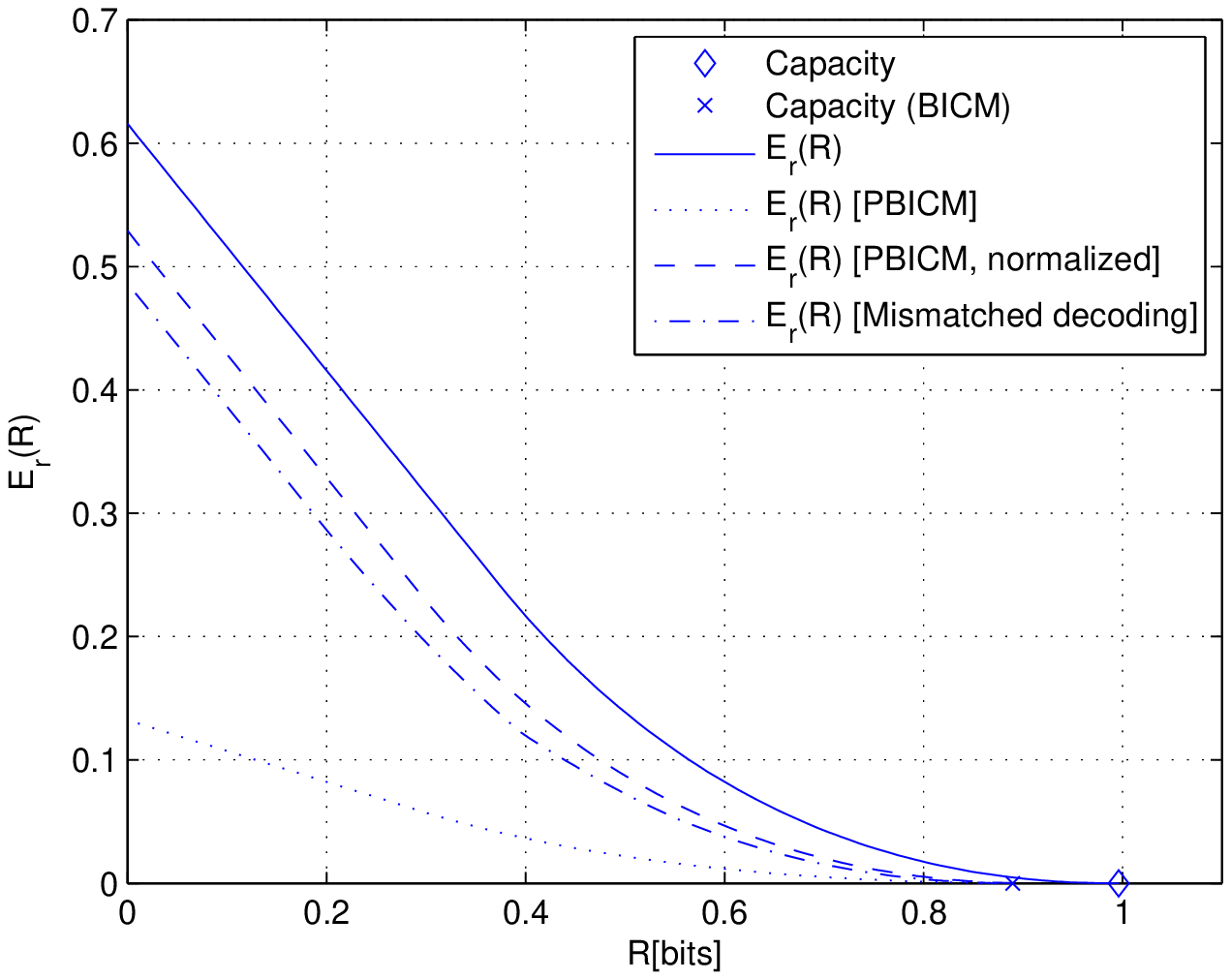}\\
  \caption{Random coding exponents over the AWGN channel with 16QAM signaling and SNR of 0dB}\label{fig:AWGN_16QAM_0dB}
\end{center}\end{figure}

\begin{figure}\begin{center}
  \includegraphics[width=5in]{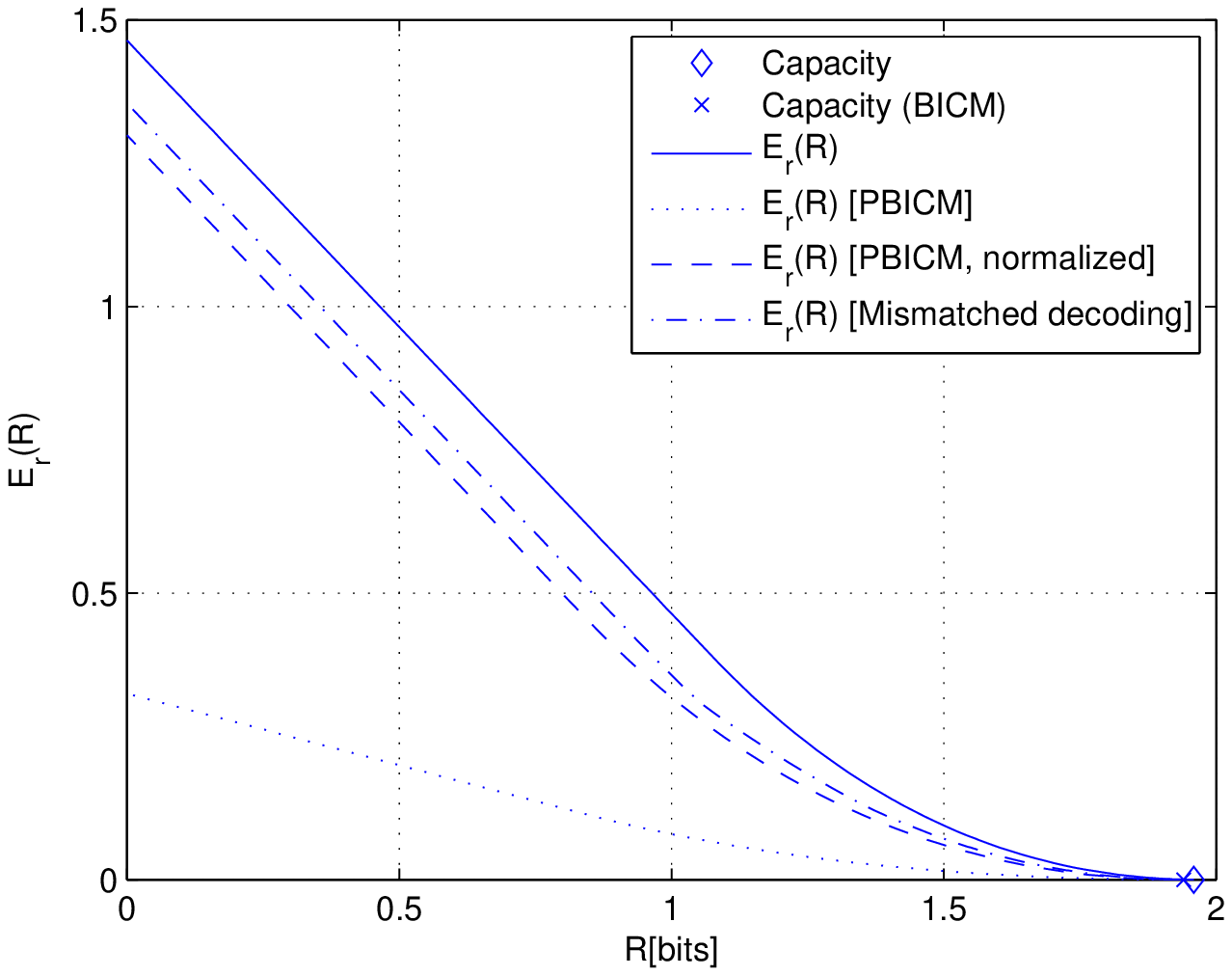}\\
  \caption{Random coding exponents over the AWGN channel with 16QAM signaling and SNR of 5dB}\label{fig:AWGN_16QAM_5dB}
\end{center}\end{figure}

On the AWGN channel it cannot be claimed that the normalized PBICM exponent outperforms the mismatched exponent, and the other way around is also not true: for 16QAM signaling and a SNR of 0dB (Fig. \ref{fig:AWGN_16QAM_0dB}) the normalized PBICM exponent was better, while for a SNR of 5dB the mismatched exponent was better (Fig. \ref{fig:AWGN_16QAM_5dB}).


\section{Discussion}\label{sec:summary}

In this paper we have presented \emph{parallel bit-interleaved coded modulation} (PBICM). The scheme is based on a finite-length interleaver and adding binary dither to the binary codewords. The scheme is shown to be equivalent to a binary memoryless channel, therefore the scheme allows easy code design and exact analysis. The scheme was analyzed from an information-theoretical viewpoint, and the capacity, error exponent and the dispersion of the PBICM scheme were calculated.

Another approach for analyzing BICM at finite block length was proposed in \cite{MartinezFabregas09_BICM_mismatched}, where BICM is thought of as a mismatched decoder. Since this BICM setting uses finite length, the random coding error exponent of the scheme can be calculated. In the previous section we have compared the error exponents of PBICM and of the mismatched decoding approach. When the two schemes have the same latency (same block length) the PBICM exponent is inferior to that of the mismatched decoding approach. However, when the complexity of the scheme is considered (or equivalently, when codeword length of the underlying code is the same), PBICM becomes comparable, and generally better over the Rayleigh fading channel.

An important merit of the PBICM scheme is that it allows an easy code design. In PBICM, one has to design a binary code for a memoryless binary channel. In recent years there have developed methods to design very efficient binary codes, such as LDPC codes \cite{RichardsonShokrollahiUrbanke_LDPC01}. When designing LDPC codes, A desired property of a binary channel is that its output will be symmetric. It appears that no matter what channel $W$ we have at hand, the resulting binary channel $\WbicmSym$ is always output-symmetric (when the output is the LLR).

Because of its simplicity and easy code design, we conclude that PBICM is an attractive practical communication scheme, which also allows exact theoretical analysis.

\newpage
Several additional notes can be made:
\begin{itemize}
  \item The analysis holds for any mapping $\mu$. Finding the mapping that yields the optimal performance at finite lengths is an open question (although Gray mapping is expected to perform well).

  \item PBICM scheme is composed of, among other things, binary dither. Such tool is used in some cases as a theoretical tool for proving achievability in some problems. In PBICM, it is an essential part of the scheme itself, and even the random capacity proof becomes impossible without it. The main role of the dither is to validate the equivalence of the PBICM scheme to a binary memoryless channel. In addition, the binary dither is the element that symmetrizes the binary channel, which makes the code design easier. This symmetrization property was also noticed by \cite{HouSMP03_ChannelAdapters} where a similar dither is used with BICM (and termed 'channel adapters'). The code design proposed in \cite{HouSMP03_ChannelAdapters} rely on the assumption of an ideal interleaver.

  \item The channel is assumed to be memoryless. This captures many interesting channels, including the AWGN channel, and the memoryless fading channel with and without state known at the receiver (ergodic fading).  For slow-fading channels, another interleaver (symbol interleaver) is required in order to transform the slowly fading channel into a fast-fading channel (cf. \cite{Caire_BICM_98}).

\end{itemize}

\appendices

\section{Approximation of the inverse Q-function}\label{app:invQapprox}
The following is a useful approximation for the inverse Q-function.
\begin{lem}
\begin{equation}\label{eqn:invQapprox}
    \lim_{\eps \ra 0} \left[\frac{(Q^{-1}(\eps))^2}{2\ln\frac{1}{\eps}}\right] = 1.
\end{equation}
\begin{proof}

We start with the well known bound on the Q function:
\begin{equation}
    \frac{1}{\sqrt{2\pi}x}\left(1 + \frac{1}{x^2}\right)e^{-\frac{x^2}{2}} \leq Q(x) \leq \frac{1}{\sqrt{2\pi}x}e^{-\frac{x^2}{2}}
\end{equation}
Dividing by the upper bound yields
\begin{equation}
    \left(1 + \frac{1}{x^2}\right) \leq \frac{Q(x)}{\frac{1}{\sqrt{2\pi}x}e^{-\frac{x^2}{2}}} \leq 1.
\end{equation}
Taking the limit $x \ra \infty$ gives
\begin{equation}
    \lim_{x \ra \infty} \frac{Q(x)}{\frac{1}{\sqrt{2\pi}x}e^{-\frac{x^2}{2}}} = 1.
\end{equation}
Since the limit exists, we may take the natural logarithm:
\begin{equation}
    \lim_{x \ra \infty} \ln \frac{Q(x)}{\frac{1}{\sqrt{2\pi}x}e^{-\frac{x^2}{2}}} = 0.
\end{equation}

\begin{equation}
    \lim_{x \ra \infty} \ln Q(x)-\ln    \frac{1}{\sqrt{2\pi}x} -\ln e^{-\frac{x^2}{2}} = 0.
\end{equation}
Since $\lim_{x\ra\infty}\ln Q(x) = -\infty$, we get
\begin{equation}
    \lim_{x \ra \infty} \frac{\ln Q(x)-\ln    \frac{1}{\sqrt{2\pi}x} -\ln e^{-\frac{x^2}{2}}}{\ln Q(x)} = 0,
\end{equation}
which leads to
\begin{equation}
    \lim_{x \ra \infty} \frac{\ln e^{-\frac{x^2}{2}}}{\ln Q(x)} = \lim_{x \ra \infty} \frac{-x^2}{2\ln Q(x)}=1.
\end{equation}
Since $\lim_{\eps\ra 0}Q^{-1}(\eps) = \infty$, we may substitute $x$ with $Q^{-1}(\eps)$, and write
\begin{equation}
    \lim_{\eps \ra 0} \frac{-(Q^{-1}(\eps))^2}{2\ln \eps} = 1,
\end{equation}
which leads to (\ref{eqn:invQapprox}).
\end{proof}
\end{lem}

\section{Big-O notation:}\label{app:notation}
As usual, $ f(n) = O(\eps_n)$ means that there exist $c>0$ and $n_0>0$ s.t. for all $n>n_0$, $ |f(n)| \leq \eps_n$ or equivalently, that
\begin{equation}\label{eqn:Onotation}
    -c \eps_n \leq f(n) \leq c \eps_n.
\end{equation}

$f_n = g_n + O(\eps_n)$ will mean that $f_n - g_n = O(\eps_n)$, which means that $f_n$ can be approximated by $g_n$, up to a factor that is not greater in absolute value than $c \cdot \eps_n$ for some constant $c$.

Sometimes we will be interested in only one of the sides in (\ref{eqn:Onotation}). For that purpose, $ f(n) \leq O(\eps_n)$ means that there exist $c>0$ and $n_0>0$ s.t. for all $n>n_0$, $ f(n) \leq c\cdot\eps_n$, and $ f(n) \geq O(\eps_n)$ will mean that there exist $c>0$ and $n_0>0$ s.t. for all $n>n_0$, $ -f(n) \leq c\cdot\eps_n$.

The different combinations of usages of the O notation are listed in the table below.
\begin{center}
\begin{tabular}{|c|c|}
  \hline
  Notation & Meaning \\ \hline
  $f_n = O(\eps_n)$ & $\exists_{c>0,n_0>0} \forall_{n>n_0} \ \ |f_n|\leq c\cdot \eps_n$ \\
  $f_n = g_n + O(\eps_n)$ & $f_n - g_n = O(\eps_n)$ \\
  $f_n \leq O(\eps_n)$ & $\exists_{c>0,n_0>0} \forall_{n>n_0} \ \ \  f_n\leq c\cdot \eps_n$ \\
  $f_n \leq g_n + O(\eps_n)$ & $f_n - g_n \leq O(\eps_n)$ \\
  $f_n \geq O(\eps_n)$ & $-f_n \leq O(\eps_n)$, or \ $\exists_{c>0,n_0>0} \forall_{n>n_0} -\!f_n\leq c\cdot \eps_n$ \\
  $f_n \geq g_n + O(\eps_n)$ & $f_n - g_n \geq O(\eps_n)$ \\
  \hline
\end{tabular}
\end{center}
Note that $f_n \leq O(\eps_n)$ with $f_n \geq O(\eps_n)$ is equivalent to $f_n = O(\eps_n)$, as expected.

\section*{Acknowledgment}

Interesting discussions with A. G. i F\`abregas are acknowledged.

\bibliographystyle{IEEEtran}
\bibliography{./bib/Master}

\begin{thebibliography}{10}
\providecommand{\url}[1]{#1}
\csname url@samestyle\endcsname
\providecommand{\newblock}{\relax}
\providecommand{\bibinfo}[2]{#2}
\providecommand{\BIBentrySTDinterwordspacing}{\spaceskip=0pt\relax}
\providecommand{\BIBentryALTinterwordstretchfactor}{4}
\providecommand{\BIBentryALTinterwordspacing}{\spaceskip=\fontdimen2\font plus
\BIBentryALTinterwordstretchfactor\fontdimen3\font minus
  \fontdimen4\font\relax}
\providecommand{\BIBforeignlanguage}[2]{{%
\expandafter\ifx\csname l@#1\endcsname\relax
\typeout{** WARNING: IEEEtran.bst: No hyphenation pattern has been}%
\typeout{** loaded for the language `#1'. Using the pattern for}%
\typeout{** the default language instead.}%
\else
\language=\csname l@#1\endcsname
\fi
#2}}
\providecommand{\BIBdecl}{\relax}
\BIBdecl

\bibitem{ZehaviBICM94}
E.~Zehavi, ``8-{PSK} trellis codes for a {R}ayleigh channel,'' \emph{IEEE
  Trans. on Communications}, vol.~40, no.~5, pp. 873--884, May 1992.

\bibitem{Caire_BICM_98}
G.~Caire, G.~Taricco, and E.~Biglieri, ``Bit-interleaved coded modulation,''
  \emph{IEEE Trans. on Information Theory}, vol.~44, no.~3, pp. 927--946, 1998.

\bibitem{GallagerInfoTheoryBook}
R.~G. Gallager, \emph{Information Theory and Reliable Communication}.\hskip 1em
  plus 0.5em minus 0.4em\relax New York, NY, USA: John Wiley \& Sons, Inc.,
  1968.

\bibitem{Strassen62_Asymptotische}
V.~Strassen, ``Asymptotische absch\"{a}tzungen in shannon's
  informationstheorie,'' \emph{Trans. Third Prague Conf. Information Theory,
  1962, Czechoslovak Academy of Sciences}, pp. 689--723.

\bibitem{PolyanskiyPV09_GaussianDispersion}
Y.~Polyanskiy, V.~Poor, and S.~Verd\'u, ``Dispersion of {Gaussian} channels,''
  in \emph{Proc. IEEE International Symposium on Information Theory}, 2009, pp.
  2204--2208.

\bibitem{PolyanskiyPVFiniteLength10}
Y.~Polyanskiy, H.~Poor, and S.~Verd\'u, ``Channel coding rate in the finite
  blocklength regime,'' \emph{IEEE Trans. on Information Theory}, vol.~56,
  no.~5, pp. 2307 --2359, May 2010.

\bibitem{WachsmannFischerHuber_MLC_99}
U.~Wachsmann, R.~F.~H. Fischer, and J.~B. Huber, ``Multilevel codes:
  Theoretical concepts and practical design rules,'' \emph{IEEE Trans. on
  Information Theory}, vol.~45, no.~5, pp. 1361--1391, 1999.

\bibitem{MartinezFabregas09_BICM_mismatched}
A.~Martinez, A.~Guill\'en~i F\`abregas, G.~Caire, and F.~Willems,
  ``Bit-interleaved coded modulation revisited: A mismatched decoding
  perspective,'' \emph{IEEE Trans. on Information Theory}, vol.~55, no.~6, pp.
  2756--2765, June 2009.

\bibitem{FoundadationsAndTrendsBICM2008}
A.~Guill\'en~i F\`abregas, A.~Martinez, and G.~Caire, ``Bit-interleaved coded
  modulation,'' \emph{Foundations and Trends in Communications and Information
  Theory}, vol.~5, no. 1-2, pp. 1--153, 2008.

\bibitem{RichardsonShokrollahiUrbanke_LDPC01}
T.~J. Richardson, M.~A. Shokrollahi, and R.~L. Urbanke, ``Design of
  capacity-approaching irregular low-density parity-check codes,'' \emph{IEEE
  Trans. on Information Theory}, vol.~47, no.~2, pp. 619--637, 2001.

\bibitem{HouSMP03_ChannelAdapters}
J.~Hou, P.~H. Siegel, L.~B. Milstein, and H.~D. Pfister, ``Capacity-approaching
  bandwidth-efficient coded modulation schemes based on low-density
  parity-check codes,'' \emph{IEEE Trans. on Information Theory}, vol.~49,
  no.~9, pp. 2141--2155, 2003.

\end{thebibliography}

\end{document}